\documentclass{article}

\usepackage{PRIMEarxiv}

\usepackage[utf8]{inputenc} 
\usepackage[T1]{fontenc}    
\usepackage{hyperref}       
\usepackage{url}            
\usepackage{booktabs}       
\usepackage{amsfonts}       
\usepackage{nicefrac}       
\usepackage{microtype}      
\usepackage{lipsum}
\usepackage{fancyhdr}       
\usepackage{graphicx}       
\graphicspath{{media/}}     
\usepackage{amsmath,amssymb}
\usepackage{todonotes}
\usepackage{comment}
\usepackage{xcolor}
\usepackage{longtable,booktabs,array}
\usepackage[numbers]{natbib}
\usepackage{authblk}
\usepackage{footmisc}
\usepackage{colortbl}
\usepackage{threeparttable,booktabs}

\title{Assistive AI for augmenting human decision-making}

\author[1,*]{Natabara M\'at\'e Gy\"ongy\"ossy}
\author[2,*]{Bern\'at T\"or\"ok}
\author[3]{Csilla Farkas}
\author[4]{Laura Lucaj}
\author[5]{Attila Menyh\'ard}
\author[1]{Krisztina Menyh\'ard-Bal\'azs}
\author[1]{Andr\'as Simonyi}
\author[4,1]{Patrick van der Smagt}
\author[2]{Zsolt Z\H{o}di}
\author[1]{Andr\'as L\H{o}rincz}

\affil[1]{Department of Artificial Intelligence, ELTE E\"otv\"os Lor\'and University, Budapest, Hungary}
\affil[ ]{\texttt{\{natabara, menyhard.balazskriszti, simonyi, lorincz\}@inf.elte.hu}}

\affil[2]{Institute of the Information Society, Ludovika University of Public Service, Budapest, Hungary}
\affil[ ]{\texttt{\{torok.bernat, zodi.zsolt\}@uni-nke.hu}}

\affil[3]{Department of Computer Science and Engineering, University of South Carolina, Columbia, South Carolina, USA}
\affil[ ]{\texttt{farkas@cse.sc.edu}}

\affil[4]{Machine Learning Research Lab, Volkswagen Group, Munich, Germany}
\affil[ ]{\texttt{\{laura.lucaj, smagt\}@argmax.ai}}

\affil[5]{Department of Civil Law, ELTE E\"otv\"os Lor\'and University, Budapest, Hungary}
\affil[ ]{\texttt{menyhard@ajk.elte.hu}}

\begin{document}
\maketitle

\begin{abstract}
Regulatory frameworks for the use of AI are emerging. However, they trail behind the fast-evolving malicious AI technologies that can quickly cause lasting societal damage. In response, we introduce a pioneering Assistive AI framework designed to enhance human decision-making capabilities. This framework aims to establish a trust network across various fields, especially within legal contexts, serving as a proactive complement to ongoing regulatory efforts. Central to our framework are the principles of privacy, accountability, and credibility.
In our methodology, the foundation of reliability of information and information sources is built upon the ability to uphold accountability, enhance security, and protect privacy.
This approach supports, filters, and potentially guides communication, thereby empowering individuals and communities to make well-informed decisions based on cutting-edge advancements in AI. Our framework uses the concept of Boards as proxies to collectively ensure that AI-assisted decisions are reliable, accountable, and in alignment with societal values and legal standards.
Through a detailed exploration of our framework, including its main components, operations, and sample use cases, the paper shows how AI can assist in the complex process of decision-making while maintaining human oversight. The proposed framework not only extends regulatory landscapes but also highlights the synergy between AI technology and human judgement, underscoring the potential of AI to serve as a vital instrument in discerning reality from fiction and thus enhancing the decision-making process. Furthermore, we provide domain-specific use cases to highlight the applicability of our framework.
\end{abstract}


\def\thefootnote{*}\footnotetext{These authors contributed equally to this work.}
\def\thefootnote{\arabic{footnote}}

\newcommand\blfootnote[1]{%
  \begingroup
  \renewcommand\thefootnote{}\footnotetext{#1}%
  \addtocounter{footnote}{-1}%
  \endgroup
}

\blfootnote{\textbf{Funding:} Our research has been supported by the Ministry of Culture and Innovation of Hungary from the National Research, Development and Innovation Fund,
financed under the TKP2021-NVA funding scheme; the "MOBOT" project (no.\ 2020-1.1.2-PIACI-KFI-2020-00115), which was implemented with the support provided from the National Research, Development and Innovation Fund of Hungary, financed under the 2020-1.1.2-PIACI-KFI funding scheme; the European Union project RRF-2.3.1-21-2022-00004 within the framework of the Artificial Intelligence National Laboratory; and the European Union’s Horizon 2020 research and innovation programme under grant agreement No 952026.}

\section{Introduction}\label{introduction}

\subsection{Motivation and AI}\label{motivation-and-ai}

Advances in Artificial intelligence (AI) reached the level that they approach or
even surpass human performance and can be used in a variety of
applications.  These advances are driven by research and development on foundation models
\citep{Bommasani2022}, such as GPT-4 coupled with DALL-E, BERT, and
related specific tools, such as Stable Diffusion, neural radiance fields,
CodeLlama.  We are in a transition phase, where the integration of AI
into society profoundly shapes the future and well-being of humanity.
The consequences of current and evolving AI-driven advancements in social,
economic, and international realms are challenging to fully comprehend;
including the intricacies of and rapid pace of the transformation unfolding
during this integration.

\begin{figure}
    \centering
    \includegraphics[width=1\textwidth]{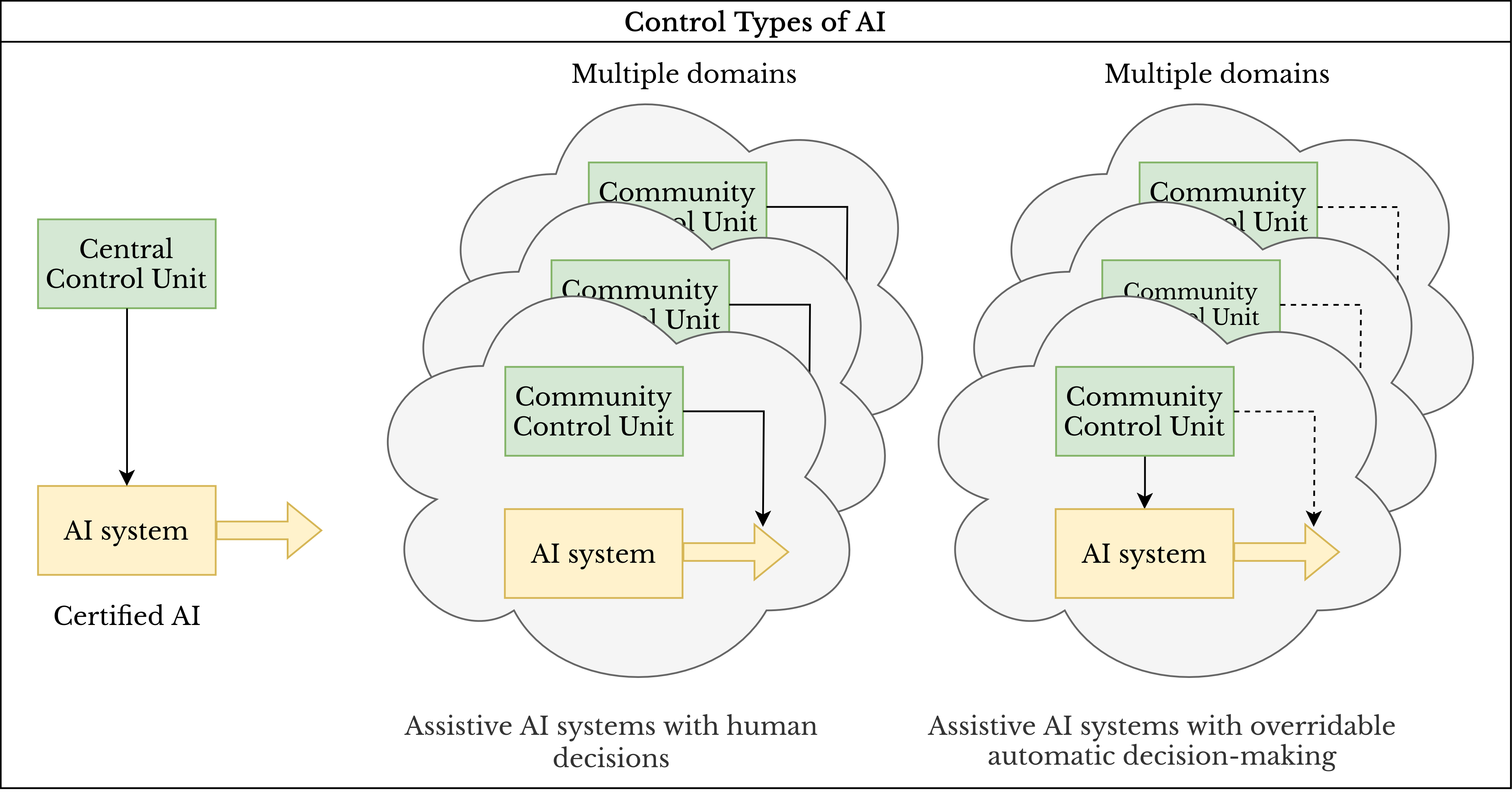}
    \caption{Certified AI systems (left) are controlled/validated by a central control unit. Domain and Community-dependent Assistive AI when used with human decisions only (middle) controls for the AI output. The overseeing unit community is controlled here. Community control units can certify AI systems as well (right) this way automated decision-making is possible, while the system still facilitates human overrides if needed.}
    \label{fig:fig_controltypes}
\end{figure}

In the long term, the emergence of superhuman intelligence will influence this
transformation. Reaching this level may be close, but neither that achievement
nor its consequences are within the scope of this paper. We are concerned about
the misuse of AI. It is already present and can create
substantial pitfalls for society. Powerful hacker tools and personalizable text
generation software have already emerged, and they don't require
advanced technical expertise. The hacker tools contain established attack tools
and are capable of automated attacks. Text generation tools can generate persuasive fake phishing texts and email accounts. In addition, social media
platforms spread fake information very quickly~\citep{Ceylan2023} .~How can we get
reliable information to make informed decisions that are in line with our goals
and values, and are congruent with the law?

The debate on AI seems to focus primarily on whether and to what extent the
technology can replace human resources in different professions or activities
\citep{Noordt2023,Laangstedt2023,Lou2023}, how to manage AI risks \citep{2023},
and how to regulate AI, see, e.g., \cite{Madiega2023}, and \cite{Tzavaras2023}
and the references to the regulations in different countries therein.
Meanwhile, there is less emphasis on how AI can help us to make
decisions on a sound basis. The reality is that, both professional services
and most platform services, AI is already helping people to make decisions
\citep{Sato2023}, while the reliability of the information technologies on which
those decisions are based is questionable and falsification of information is
possible.

Some decisions can be made by AI itself.  For example, consider the decisions on acceleration and
slowing down for self-driving cars.  Such AI decisions can be achieved given appropriate testing and a social consensus on the associated risks. However, when
weighing and prioritizing social values, or when the decision is subject
to the scrutiny and interpretation of legal norms, the situation is different.

A model is needed that strikes an optimal balance between efficiency and
protection of society's values even in the absence of human control.  For example, it is unrealistic to anticipate sufficient human capacity to filter out the
flood of automatically generated fake news.

We introduce the concept of a "Board". A Board has functions that include leadership, decision-making, and oversight responsibilities. To describe the network of decision-making units, we also talk about "consultants", who give advice. We borrow the term "auditor" from the financial sector for experts whose tasks include independent evaluation, verification, and oversight to ensure compliance with standards, regulations, or best practices. Examples include Quality Assurance Inspectors, Compliance Officers, and IT Auditors, among many others. An auditor is typically an independent entity.   

\begin{figure}
\centering
\includegraphics[width=0.7\textwidth]{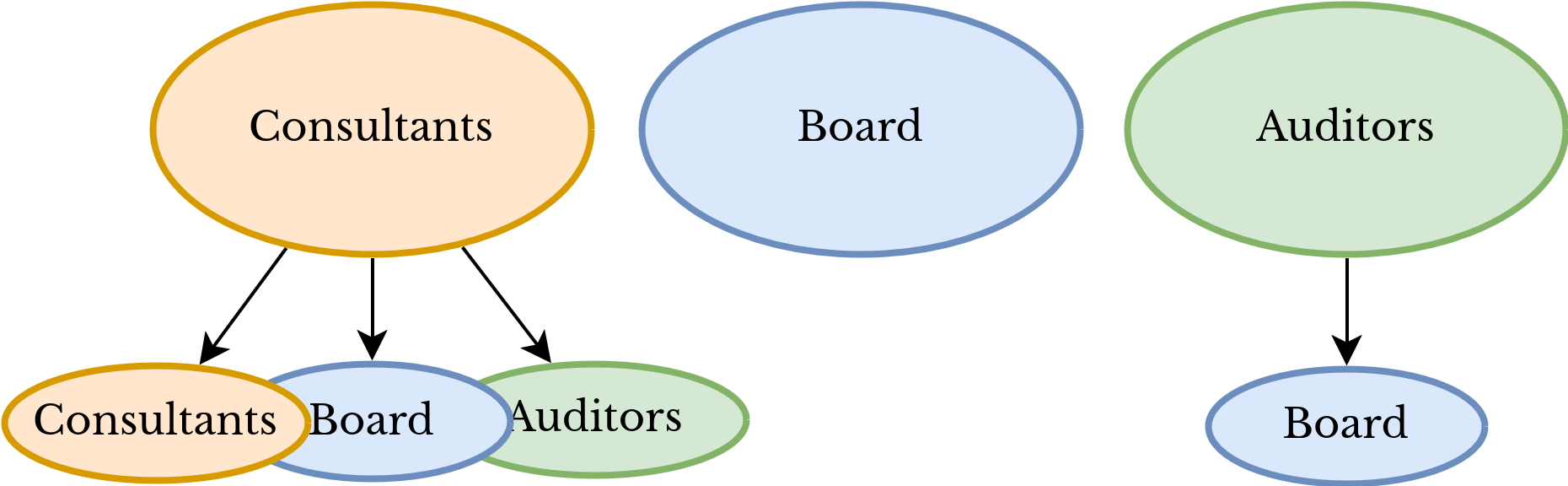}
\caption{The Board can be connected to consultants and auditors. Each unit, e.g., a consulting company may have a Board and may have associated external consultants and auditors, too.}
\label{fig:board}
\end{figure}

Our concepts are sketched in Fig.~\ref{fig:board}. We elaborate on two examples, the editorial board of a journal and the relationship between the Food and Drug Administration and the Internal Review Boards of clinics in a later section. 

Automated decision-making should be possible where legality is verified
through regulation. However, where the law mediates the social
evaluation of the case and the judgments of the courts are derived from
the consideration of the relevant values of the case, the replacement of
human resources (the judge) by artificial intelligence is less
supported, at least in the current structure of our societies. Still,
the use of artificial intelligence to assist human decision-making is a
practical and realistic idea if an appropriate framework can be
designed.

\subsection{Why Assistive AI?}\label{why-assistive-ai}

In this paper, we aim to introduce a concrete model for exploiting the positive
application opportunities inherent in AI to those individuals and their
communities that are looking for reliable guides in the increasingly
the complex digital reality that surrounds us.   The concept that we present here complements recent developments related to the regulation of AI (see Figure \ref{fig:fig_controltypes}).  However, we are offering a
different approach by supporting safe, reliable, and accountable AI applications.  The initiatives of the
European Union -- the AI Act and the AI Liability Directive (AILD) -- can be
considered as particularly remarkable regulatory developments worth
supplementing with application models in a complementary approach.
In addition to the AI Act (which focuses on the common risks associated
with the use of AI systems), and the AILD (which settles the liability of potential damage
caused by AI), the present concept outlines a possible AI model that can
serve as a concrete practical solution to some challenging issues.

Regulations, especially certain requirements of the future AI Act, are
important starting points for our model, but our focus is on moving
forward and supplement the regulatory approach as follows:
\begin{itemize}
\item
  While the regulation focuses on the common negative aspects of AI
  systems (their risks or damages) and their management, we outline a
  special application model of AI, with which AI itself can become one
  of our important supports in clarifying the reality around us.
\item
  While in the regulation, AI is the object of control (that is, we
  bring AI under control), in our model, AI is the means of control
  (which helps us in bringing other processes under control).
\item
  While for the regulation, AI is a product whose development and application must be guided by requirements, for us, AI is a special
  instrument to examine the enforcement of requirements on other
  products, outputs, etc.\ (in our terminology below: communications).
\item
  While the regulation tries to protect the individual against certain
  applications of AI systems along the lines of general
  socio-legislative goals, we try to give them an AI-based tool to
  protect themselves according to their own goals and aspects.
\end{itemize}

\rowcolors{2}{gray!25}{white}
\begin{table}[h]
\centering
\caption{Comparison between Regulatory Approach and Approach of Assistive AI}\label{tab:comp}
\begin{tabular}{
  >{\raggedright\arraybackslash}p{0.49\linewidth}
  >{\raggedright\arraybackslash}p{0.49\linewidth}
}
\toprule
Regulatory Approach & Approach of Assistive Artificial Intelligence \\
\midrule
In the focus: risk caused by AI (generally) & In the focus: Assistive Artificial Intelligence as a special opportunity \\
\addlinespace[4pt] 
AI is the object of regulation (something to be controlled) & AI is the tool that helps us to control (something to rely on during decision making) \\
\addlinespace[4pt] 
AI is a product that we put requirements on & AI is an instrument to call attention to and check requirements on other products/outputs (communications) \\
\addlinespace[4pt] 
Regulating AI as technology generally & Modelling Assistive Artificial Intelligence as a special application of AI \\
\addlinespace[4pt] 
Societal goals are in the focus & Individual/community goals are in the focus (with social goals in the background) \\
\addlinespace[4pt] 
Protecting passive individuals & Activating individuals and communities to protect themselves \\
\bottomrule
\end{tabular}
\end{table}

As outlined in Table~\ref{tab:comp}, the key is the development of a methodology that connects humans and AI that involves the identification and weighing of conflicting societal values using algorithms. The goal is to maximize human control and governance while accounting for the enormous flood of information by enabling automatic processes considered safe by accountable entities. The right combination of anonymity and accountability, as well as the credits of the authors/makers of products, supports the goal. While the system is derived from higher-order principles and rules, it can serve communities at different levels having their own goals. In all cases, the role of AI is to assist the community and the role of our proposed framework is to maintain human control while taking advantage of Assistive AI.

The abstract Assistive AI framework proposed in our work aims at new knowledge management that can support decisions at diverse levels, while also providing estimations, sometimes guarantees on the correctness of the information.

\subsection{Central concepts}

There are a few critical concepts and components that we consider necessary for the efficient and correct functioning of Assistive AI. 
A brief overview of these concepts is given below. 

\subsubsection{Anonymity and accountability}

In an era rife with misinformation, we believe it is crucial to have systems that ensure authors are accountable for their content to foster trust. However, scenarios exist where authors might need to stay anonymous, such as when revealing information ethically necessary but contractually restricted. For example, if a breach of the law is observed but a contract prevents the disclosure of this information. Conversely, anonymous disclosures can inadvertently cause harm or spread falsehoods, raising liability issues. A solution involves algorithms that preserve anonymity until a legitimate authority deems it necessary to reveal the author's identity for accountability. We shall elaborate on this critical issue in Section~\ref{accountability}.

\subsubsection{Credit and credibility}

Similarly to anonymity and accountability, the reputation or credit of an author could convey the significance of the author's communications. There are numerous approaches to grasp the real meaning behind credibility. Hovland, et al.\ define it from the communication science perspective, where credibility refers to the property of a source based on the recipient's acceptance of the source's message~\citep{hovland1953communication}. Other definitions include credibility as a perceived value by the recipient or the relevance and truth value of the conveyed information~\citep{rieh2007credibility}. Numerous studies point out that in modern media, human-computer interactions, and online communication credibility is closely related to relevance, communication quality, truthfulness, and eventually the trust of the recipient~\citep{10.1111/jcc4.12084,rieh2007credibility,viviani2017credibilityinsm,wathen2002factorsofwebcredibility}.

In this work, we identify two possible sources of communication and interaction:
a) \emph{Authors} and a \emph{Board}.  
 \emph{Authors} of communications belong to a given community. A \emph{Board} oversees a given community by designing and controlling its internal policy (for details see Section~\ref{sec:boards}). \emph{Authors} and \emph{Boards} have credit scores to denote their credibility.  While we believe that the exact method of author credit calculation is domain and community-dependent, we argue that Assistive AI systems, through network and communication analysis, are capable of providing automatized assistance in credibility estimation. This is essential, as communities empowered by such Assistive Artificial Intelligence could effectively estimate the relevance and truthfulness of communications and authors; even if a human evaluation would not be viable due to the huge number of communications~\citep{MEEL2020112986}. To distinguish our usage of the word credit from the financial usage and highlight the community and \emph{domain-dependency} we will refer to this credibility of the authors as \emph{d-credit} for short, while keeping the \emph{credit score} expressions for the estimated level of d-credit.

\subsubsection{Role of Boards}
\label{sec:boards}
\begin{figure}
    \centering
    \includegraphics[width=0.9\textwidth]{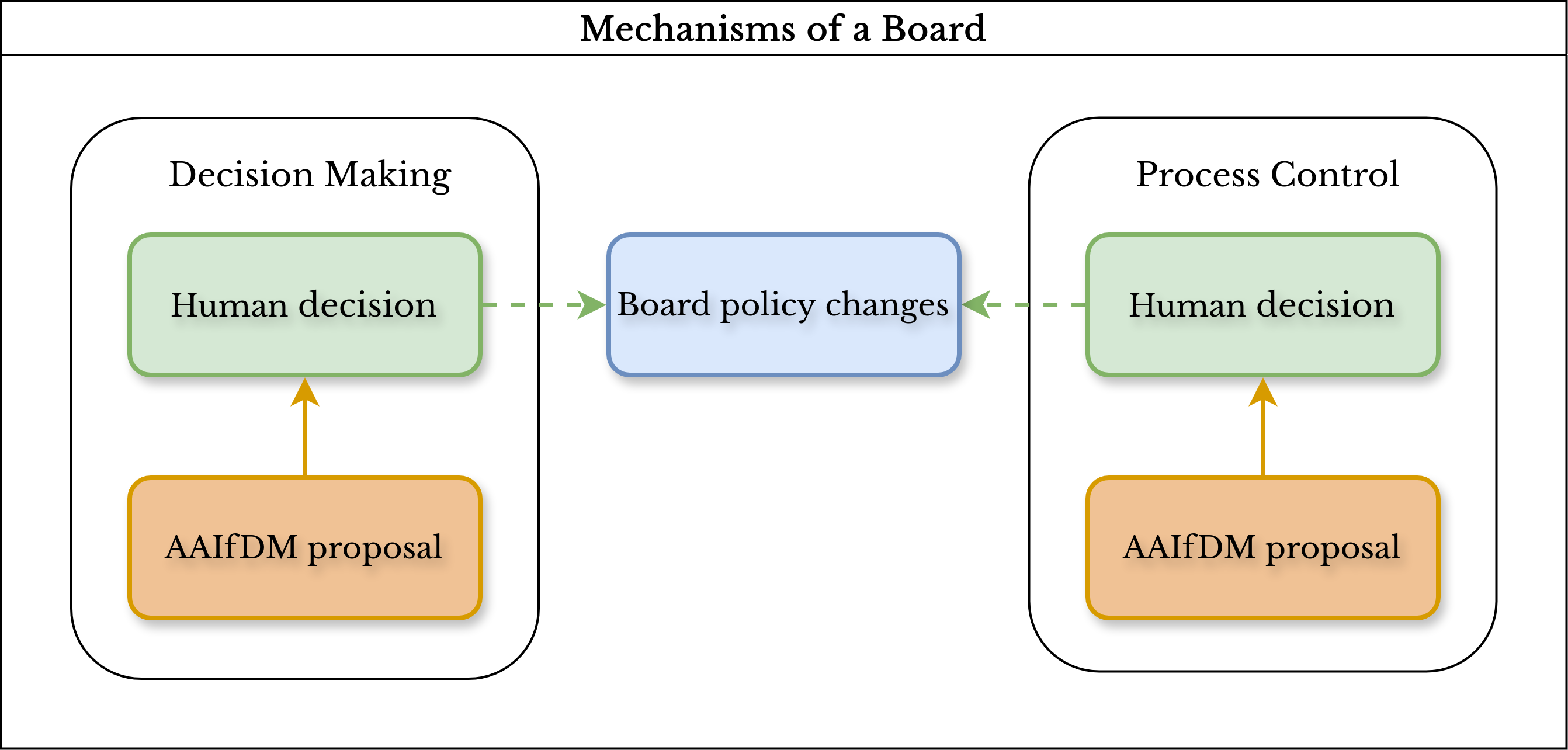}
    \caption{Boards typically assume dual functions. First (left), they decide on policy changes, especially if the environment changes due to AI developments. These changes affect everyone in the community, including other boards. Second (right), they perform process control, i.e., review and control policy changes of the boards in their community, ensuring these changes follow the community's values and legal standards. The Board can change, add to, or clarify the policy changes of the Boards of its community. In all cases, Assistive Artificial Intelligence can support the decision-making procedure. }
    \label{fig:board_mechanism}
\end{figure}

Through the supervision of legal compliance, the State ensures that the concerned entities affected by the legal act in the course of exercising their rights and obligations, or their operations shall comply with the provisions of the legislation in force.

Our concept, illustrated by Figure  \ref{fig:board_mechanism}, is based on the fact that in communication data must be authentic, up-to-date, real, and legitimate, i.e.,\ not only the operation of an organization or a community must be legitimate, but also the legitimacy of the data/d-credit must be ensured in the process, on the reason that the two are closely in line and interact back and forth.

The authority in charge of supervision of legal compliance, in our case the Board, has the right to indicate or authorize a correction based on findings.

If the organization or community fails to remedy a situation that is in breach of (law), to commence the procedure may be requested for as long as the situation persists and shall continue until the lawful situation or lawful operation is restored.

The Board shall take its action depending on the circumstances and severity of those circumstances.

\subsection{Main contributions}\label{main-contributions}
Our main contributions are as follows:
\begin{itemize}
    \item We put forth a novel Assistive Artificial Intelligence framework for communities. The basic unit of the community is the Board which must be accountable by law. Boards are partially ordered and hierarchically extendable.
    \item We show how cutting-edge Assistive Artificial Intelligence tools may complement and extend regulatory environments. Our proposed Assistive AI assesses the reliability of communications, serving as a proactive tool that can also evaluate the credibility of information sources. It is also fast and able to keep up with the revolutionary AI developments, supporting decision-making, and securing community interests. 
    \item Assistive Artificial Intelligence framework combines anonymity, accountability, reliability, and legal compliance, decreasing the risk in AI governance.
    \item At the heart of the Assistive Artificial Intelligence framework is the collaboration between humans and AI, with automation capabilities but under human control.
    \item Assistive Artificial Intelligence is a high-level concept at present involving cutting-edge AI and cybersecurity methods. Its physical realization is feasible as the components are available. Thus, incremental, step-by-step construction and implementation of the Assistive Artificial Intelligence system is feasible. We review the testing methods required by Assistive Artificial Intelligence for the AI tools to be included. 
    \item We provide 4 examples, such as recommendation on advertisements of drugs or health services, opinion about the credibility of communication on academic subjects, financial product recommendation, and financial credit scoring.
\end{itemize}

\subsection{Structure of the paper}\label{structure-of-the-paper}

The paper is organized as follows. We start with a section on related works~\ref{related-work}. This section briefly reviews, e.g., AI-based recommendations, and community and public sector decision-making. We list the critical components that we need for our work and provide the references. Section~\ref{ss:background} is about components, such as the Board concept, questions related to security and privacy, anonymity and accountability, and ethical issues. The Assistive Artificial Intelligence Framework is treated in the next section (Sect.~\ref{aai-framework}). Sample use cases can be found in Sect.~\ref{sample-use-cases}. We discuss our approach in Sect.~\ref{discussion-and-future-work}. Conclusions are drawn in Sect.~\ref{conclusion}. The appendix (Sect.~\ref{appendix}) contains additional AI-related details about the financial use case (Sect.~\ref{appendix:financial-use-case-ai-background}), detailed consideration of recommendation and advertisements for drugs and health services (Sect.~\ref{recommendation-on-advertisements-of-drugs-or-health-services-ai-background}), and our opinion on the credibility of communication of academic subjects (Sect.~\ref{opinion-about-the-credibility-of-communication-on-academic-subjects-1}). 

\section{Related work}\label{related-work}

Our focus lies in using Assistive Artificial Intelligence to enhance decision-making processes, which can apply to individual choices or decision-making within groups and public entities. On an individual level, AI-driven personalization, like the kind seen in services offered by Netflix, Amazon, Google, and numerous other online platforms, provides tailored recommendations that aid users in decision-making. Similarly, AI technologies may offer valuable support to both patients and healthcare professionals by delivering personalized healthcare recommendations. Additionally, this customized approach is increasingly adopted in the financial sector, guiding individuals in managing their finances and making investment choices more effectively, but also raising ethical challenges, and risks due to the lack of transparency and the possibility of generated fake information (see, e.g., \citet{khan2024chatgpt} and the references therein).

Community and public sector decision-making processes are closer to our theme. In this case, resource allocation, urban planning, and public safety could serve smart cities \citep{im2023exploring}. Applications may arise in managing public health crises, including pandemic response and vaccine distribution strategies \citep{syrowatka2021leveraging}. Environmental management is another field. Here, AI may provide advice in environmental protection and sustainability efforts, including climate modelling and conservation strategies, see, e.g., \citep{taghikhah2022artificial, silvestro2022improving}.
ELSI (Ethical, Legal, and Social Implications) can also take advantage of Assisitive AI, e.g., in discussions on (a)~biases in AI algorithms and their impacts on decision-making for diverse populations, (b)~privacy and security issues, (c)~regulation and governance, and (d)~specific instances where AI has been successfully implemented.
New AI technologies and methods will significantly improve decision-making capabilities in different sectors. Tools that enable the combination of artificial and human intelligence to optimize decision-making processes are flourishing. AI can assist by gathering, summarizing, and collating information across data landscapes and the trend is propelled by advances in Natural Language Processing (NLP) and Understanding (NLU). Beyond the different views that frequently appear on websites \citep{meissner2023}, papers in scientific journals started to study AI approach to ethical decision making, e.g., in human resource management \citep{rodgers2023artificial}, decision support in patient care \citep{fonseca2024embracing}, and also considering the effect of superhuman intelligence on human decision making \citep{shin2023superhuman}, to list only a few.

We aim to sketch an architecture that can promote information filtering and aid human decision-making. There are critical components that should be dealt with and incorporated into such architectures but are beyond the scope of our paper. We list some of those here:
\begin{enumerate}
    \item AI methods should provide explanations for us. This is the target of Explainable AI: the outputs of the AI systems should be transparent and understandable to us at the level we want it. Algorithmic explainability should also cover legal argumentation~\citep{10.1145/3236386.3241340, Arrieta2020, DBLP:journals/corr/abs-2006-00093, zHodi2022algorithmic, SAEED2023110273}. 
    \item The development of trustworthy AI systems focuses on techniques that yield outputs which are dependable, ethically sound, and clear. These initiatives aim to embed mechanisms for ensuring accountability, safeguarding privacy, and bolstering security, ensuring that the AI's decisions are defensible and in harmony with the broader societal norms  \citep{Hagendorff_2020,10.1145/3555803}. There are various public initiatives in this area\footnote{See European Ethics Guidelines for Trustworthy AI \citep{ai2019high},  Ethical and Trustworthy Artificial and Machine Intelligence \citep{etami}, and the NIST Trustworthy and Responsible AI \citep{NIST-trustworthy} as notable examples.}. 
    \item Robustness, Resilience, and Sustainability in AI Systems focuses on preventing adversarial attacks, and data quality issues, and aims at making AI systems more resistant to errors and changes over time \citep{NEURIPS2019_e2c420d9,tocchetti2022ai}.
    \item Sociotechnical AI Systems is dealing with the interplay between AI technology and social systems with specific emphasis on social dynamic and human values \citep{sartoriSociotechnicalPerspectiveFuture2022,weidinger2023sociotechnical}.
\end{enumerate}

In their recent work, Bao et al. review the methods of human-AI decision-making \citep{bao2023synergy} and categorize these methods as follows: 
\begin{itemize}
    \item A body of papers deals with the identification of AI affordances in decision-making, such as automated information collection and updating, information processing and analysis, predicting, forecasting, assistance in decision-making, and AI explanations. 
    \item Another cluster is about human-AI synergy patterns in decision-making. This cluster includes methods that are (a) centred on AI algorithms, such as natural language processing, machine learning, and so on, (b) human-centred aiming at shared mental models. Explainable AI belongs to this cluster, and (c) human--AI synergies that emphasize the collaboration between the partners \citep{bansal2019updates}.
\end{itemize}

Our research delves into the legal dimensions of collective and community-based decision-making, emphasizing the importance of retaining human oversight in AI-assisted decision processes. While our study does not directly investigate collective decision-making, the advancements in this area are pivotal to the framework we propose, and a rapid evolution of the field can be seen. For those interested in exploring recent developments, we recommend consulting an arXiv publication on the "Equitable AI" research roundtable \citep{smithloud2023equitable} and the ACM conference series on Fairness, Accountability, and Transparency \citep{facct} that has a five-year history and brings together researchers and practitioners interested in fairness, accountability, and transparency in socio-technical systems.

\section{Background}\label{ss:background}

Before presenting the concept of Boards and discussing questions related to security and privacy, the method of accountable anonymity, and ethical issues, we examine the basic concepts of law and how artificial intelligence can be used in this field.

\subsection{Possible paths for artificial intelligence in the field of law}
\label{possible-paths-for-artificial-intelligence-in-the-field-of-law}

Law transforms plural values of the society into the binary code of right and wrong \citep{habermas1984theory}: The law involves social evaluation, reflected in the court's judgement in each individual case \citep{wilburg1950entwicklung}. 

Decisions in social conflicts take into account the specific contexts and are influenced by a careful balancing of all the interests involved. The leading guideline is equal treatment of equal situations, the desired model of judicial adjudication. 

Judgements frequently involve prioritizing competing values, such as freedom of speech versus human dignity. Consequently, any algorithm designed to support the enforcement of law must, to some extent, be capable of performing social evaluations.

Regulation, contrasted with law, is supposed to prescribe clear-cut rules for social
coordination and does not necessarily reflect values. Instead, it implements the direct
policy of the regulatory power. Algorithmic legal decision-making by artificial intelligence seems plausible in the context of regulation, as the task is to "match" the facts of the case with the content of the norm in the legislation~\citep{lai2023large,cui2023chatlaw}. The question, however, is how and to what extent this is possible when decision-making is also about the social perception of the issue.

Assume that the relevant facts, values and even political weighting can be established from previous judgements. In this case, algorithms can estimate and rank possible outcomes, provided that the relevance of social values is given, or at least limited.

Such cases certainly exist. In contract law, for example, the values reflected in legislation and judgments include private autonomy, which varies based on the parties' informed consent and freedom during the contracting process; structural inequality in bargaining power, as seen in consumer protection measures; protection of reliance on the other party's statements and conduct, addressing the extent of interest protection and the permissible limits of liability for contract breaches; the objective assessment of the intended exchange, evaluating the balance of rights and obligations between parties; and the principle of responsibility and self-reliance, emphasizing the fundamental moral value of keeping promises \citep{bydlinski2011methodological, gilead2013proportional}.

In tort cases, the relevant social values include: the potential number of plaintiffs; the necessity for an increased duty of care, if established; the closeness and particular relationships between parties; the level of danger posed; the degree of dependence; the obviousness and actual knowledge of the risk; the clear definition of the case's boundaries; the distinction between negligence and deliberate actions; and the significance of the financial interests involved for the parties \citep{koziol2010grundfragen}. It is an open question when and how artificial intelligence can help decisions, other than to collect similar cases, if any.

To have an insight, we sketch a most difficult category of the diverse cases from the point of view of the protection of social values. It is the award of non-pecuniary damages, which is the primary means of protecting and enforcing fundamental rights in private law, including the prioritisation of conflicting fundamental rights. It is also demanded for courts to be consistent in the amounts awarded. The case is extremely hard for machine evaluation as in adjudicating such cases, relevant decision-making factors include the nature and severity of the injury, the personal circumstances of the victim, and the injury's consequences. When determining the compensation amount, several factors are considered, among others, including the victim's age; the extent to which the injury reduces the victim's ability to work; the impact on specific careers such as those of athletes and musicians; the severity of the personal injury; the loss of the chance to have children or start a family; the impairment of learning and educational opportunities; the presence of aesthetic damage; the occurrence of serious mental illness or personality disorders; the reduction in quality of life; whether the victim's condition serves as a deterrent or alarming factor to others; and the effect of the injury or illness on the victim's social relationships. 

Still, in such complex cases, quantitative criteria may exist enabling estimations using algorithms if the relevant facts of the case can be identified. Then the gravity of those relevant values can also be estimated via quantitative criteria based on the outcomes of previous judgements~\citep{wu-etal-2023-precedent}. With the current scientific advancements, one may expect that such models and methods will be available in the not-to-remote future for assessing fundamental rights and suggesting priorities among them.

\subsection{The concept of Boards and related entities}

Boards, as mentioned in the Introduction, have functions that include leadership, decision-making, and oversight responsibilities, among others. Note that the word ‘Board’ can be used in different senses. To disambiguate what we mean, we provide two examples. We also describe the concepts of Consultants and Auditors (sketched in Fig.~\ref{fig:board}) through the example of the Editorial Board of the journal and the other example, the relation between the Food and Drug Administration and the Internal Review Boards of clinics. 

\subsubsection{Example 1: The editorial Board of a journal}

\textbf{Reviewers} of the Board are much like \textbf{consultants}. They offer their expertise and knowledge to assess the quality, accuracy, and relevance of the manuscript including the relevance for the intended readership. They provide constructive feedback, may suggest improvements, and provide recommendations on whether a manuscript should be accepted, revised, or rejected. Their role is advisory and supportive, aimed at enhancing the quality of the published work and thus, the venue.

Just as consultants are often hired for their specialized knowledge in a particular field, reviewers are selected based on their expertise in the subject matter of the manuscript. They apply their understanding and evaluate the contribution of the paper to the field. 

The review process is like many consulting arrangements, e.g., the work is conducted confidentially, it respects the integrity and proprietary nature of the authors' work. 

\textbf{Readers} of the publications are similar to \textbf{auditors}. Readers, akin to auditors, play a role in the post-publication phase, where they critically evaluate and interpret the research findings. They can assess the validity and applicability of the research in their work or further studies, thus acting as a form of quality control after publication. 

While auditors might provide feedback or raise concerns about the practices of an organization, readers can engage in scholarly discourse by citing, discussing, or critiquing published work in other forums, conferences, or subsequent publications. This feedback loop can influence future research directions and the reputation of the journal and its contributors. 

Auditors ensure that organizations adhere to standards and are accountable to stakeholders, and readers ensure that scientific research remains transparent, reproducible, and accountable to the academic community and society at large. 

Among other things, the role of the Editorial Board is to maintain and enhance the quality, integrity, and reputation of the journal. The board establishes the journal's editorial policies, including ethical guidelines, submission criteria, and the scope of the journal. These policies aim to maintain the journal's quality and integrity. The board contributes to the strategic direction of the journal, including identifying new areas for publication, expanding the journal's scope, and integrating emerging scientific trends and disciplines. The board ensures that all activities of the journal adhere to ethical standards, handling allegations of misconduct (e.g., plagiarism, data falsification) and implementing corrective actions when necessary. 

\subsubsection{Example 2: Food and Drug Administration and the Internal Review Board of a clinic}

The Food and Drug Administration (FDA), the Internal Review Boards (IRBs) and researchers of a clinic make another example of our multilevel structure.  

The IRB is a board that has the authority to approve, require modifications in, or disapprove research studies based on their assessment of risks and benefits to participants, the informed consent process, and the protection of participants' rights and privacy.  

The FDA, on the other hand, is another large unit and IRB and FDA are two critical components of a broader framework that governs clinical research in the USA: (a) FDA focuses on regulatory oversight for the safety and efficacy of products entering the market and (b) IRBs focuses on the ethical conduct of research and protection of human subjects. Now, the FDA has several advisory committees, including the FDA Science Board (which is not a board in our nomenclature as it does not make decisions) and members of these committees can be seen as consultants who offer guidance, recommendations and professional opinions for the FDA to make informed decisions. The Office of the Commissioner of the FDA, on the other hand, is similar to the Editorial Board of a journal even though there are fundamental differences in their roles, responsibilities, and the contexts in which they operate. However, there are aspects of their functions that can be seen as somewhat analogous, particularly in terms of leadership, decision-making, and oversight responsibilities. 

\subsection{Relevant questions related to security and privacy}\label{security-privacy-and-accountability}

\subsubsection{AI testing safety and monitoring}\label{ai-testing-safety-and-monitoring}

Currently, many issues have been exposed by academia and investigative
journalists around the absence of proper oversight mechanisms internally to test
AI systems before deployment, as well as external mechanisms, such as
regulations, certification, and third-party auditing to enforce and mandate
better practices for the companies developing such systems
\citep{Lucaj2023,Cobbe2023,Wieringa2020,Raji2022}. One of the most pressing
issues seem to remain in the monitoring and testing of AI systems, which carries
enormous costs as well as implementation issues \citep{Bennett2000}.

Many AI systems could cause significant harm. This includes foundation models, which
have been put on the market and subsequently have been found to need better mitigation
strategies to keep up with their rapidly increasing performance that can be used
for malicious intent \citep{Shoker2023,Glaese2022}. Currently AI developers
seem to have issues with scaling safety operations, which results in many cases of
bypassing the safety filters of systems such as ChatGPT or Stability AI
\citep{Shoker2023}. To prevent such harm from occurring, dedicated and strong
forms of transparency from foundation model developers are necessary
\citep{Bommasani2023}.

Testing of AI has not reached the maturity of testing in
software engineering, where testing is seen as a fundamental practice to develop
well-functioning software \citep{Brundage2020}. It is key
to learn from the software engineering field. It can guide how to
navigate the technical challenges that arise as organizations have to develop
large-scale solutions, whilst keeping the systems robust and secure by enabling
the adoption of quality assurance practices \citep{Amershi2019}. The maintenance
practices developed in the field can provide guidance for AI practitioners,
as often new technologies are deployed without the necessary tools and processes
throughout the lifecycle that allow the analysis of the potential consequences
once the software has to be changed \citep{Bennett2000}. Indeed, the whole concept of the life cycle is not yet established for AI systems.

Consequently, research struggles to keep up with the understanding and underpinning
on the maintenance and evolution practices that can be scaled in the deployment
on the market \citep{Bennett2000}. An interesting approach that can be adopted
by AI developers is the practice of analyzing diagnostics-based customer
application behaviour, prioritizing bugs, estimating failure rates, and
understanding performance regressions by involving data scientists in the development and evaluation
\citep{Kim2016}. Practices such as bug bounties, audit trails, incident
analysis and red team exercises conducted in software engineering can provide
concrete methods to enable AI developers to assess the performance of the system
as well as enabling to gain trustworthiness \citep{Avin2021}. One of the biggest
challenges are to address unknown concerns by analyzing the domain of deployment
and the limitations or risks that could be potentially exploited by actors with
malicious intentions \citep{Brundage2020}. Currently, this issue has been exposed
by the many cases of investigative journalism delineating how users manage to
bypass the safety filters of, e.g.,  ChatGPT and Stability AI.

\subsubsection{Potential practices}\label{potential-practices}

Red team exercises have been deployed in the software engineering domain to
address the potential misuse of the systems as well as unveil its potential
vulnerabilities \citep{Avin2021}. Audit trails are another important practice
that provides an overview of the important processes that determined the
development of a system as well as keeping track of the incidents and failures
emerging during the development phase
\citep{Shneiderman2020,Avin2021,Brundage2020}.

Bias and safety bounties are an established practice consisting of the
financial rewarding of security experts who need to expose potential
vulnerabilities of the system before deploying it in the market phase
\citep{Avin2021,Brundage2020}.

Nonetheless, the development of AI systems poses new challenges for
practitioners as opposed to software application domains. For instance,
the data management practices are much more complex; model customisation
and reuse require different skills from the software domain; and the AI
components are more complex to handle as distinct phases as they all
intertwine with another much more than is customary in traditional software approaches
\citep{Amershi2019,Hua2020}.

\subsubsection{Privacy preservation}\label{privacy-preservation}

Protecting the individual's private information has been studied extensively in
statistical databases (see \citealt{Denning1983,Farkas2002} for an overview).
Statistical inference control aims to protect private data while allowing the
computation of useful statistics. Inference control techniques are based on
limiting the users' statistical queries, anonymizing and suppressing data, and
introducing noise in the data in a way that the modified data still lead to
useful statistics, but the individuals' records are not real. These techniques
are still used to provide privacy protection in large datasets used for
statistical purposes, such as the United States Census Bureau.

With the advent of large-scale data analytics to support data mining
applications, the need to provide privacy protection that is suitable for these
applications surfaced. In 1998, the concept of k-anonymity was proposed by~\citealt{Samarati_k_anonimity_updated}. The promise of k-anonymity is that an individual's personal
data cannot be distinguished from at least k-1 other individuals' data. During
the early 2000s, researchers proposed methods to strengthen the k-anonymity
model by addressing attacks based on the homogeneity of the data and the attackers'
background knowledge (see \citealt{Sweeney2002, Machanavajjhala2007,Kabir2010}
for representative examples). While there are well-documented limitations of the
k-anonymity techniques, they represent a practical approach to express and
enforce privacy requirements \citep{AyalaRivera2014}. These works also formed
the foundation of further studies on privacy-preserving data mining in a variety
of context
\citep{Vaidya2002,Vaidya2003,Kantarcioglu2002,Siraj2019,Arumugam2016,Aldeen2015,Wahab2014}.
A fundamental aspect of all proposed approaches is the trade-off between
the accuracy of the data mining application and the level of privacy protection.
There are two main approaches to introducing noise in the data mining application:
1) introduce noise in the data set before the mining application and 2)
introduce noise in the learned model. In each case, the tradeoff is to achieve a
useful result while guaranteeing a privacy protection level.

Differential privacy \citep{Gaboardi2018,Niculaescu2018,Dwork2014} provides a
formal measurement of the level of privacy. Intuitively, complete privacy is
achieved if the learned model from the original dataset is the same as the model
learned by removing any of the individual's data. That is, the dataset does not
reveal any specific information about an individual. The amount of noise
introduced in the dataset is dependent on the privacy sensitivity of the
individual's personal data and the usage of the data in statistical analysis.
Local differential privacy is achieved by the individuals adding noise to the
dataset containing their personal data. The advantage of this approach is that each
individual modifies his/her data, therefore no one else will know the real
personal data. However, this approach may result in too much noise and may
reduce the accuracy of the application. Global differential privacy is achieved
by adding noise to the output of the system. The advantage of this approach is
a high level of accuracy at the cost of trust in the database curator.

Current research extends the concept of privacy preservation to the realm of
Artificial Intelligence (AI) applications (for representative examples, see
\citet{Ryffel2018,Chen2022a,Neumann2023}). In
addition to the classical definition of privacy needs, these works address
issues of distributed and federated computing, lack of trust in the
participants, and a variety of learning domains. Promising technologies, such as
homomorphic encryption \citep{Guo2023}, secure multiparty encryption
\citep{Prabhakaran2013}, and remote computation \citep{Ziller2021} being
developed to support practical and comprehensive privacy-preserving AI
applications.

\subsection{Anonymity with accountability }\label{accountability}

Accountability is the ability to hold active entities responsible for
their actions. A system may provide accountability at multiple levels.
For example, accountability may be enforced on a group instead of a
particular individual or enforced within a virtual environment without
addressing consequences in the real world. Enforcing accountability
requires some form of identity management. In digital systems, strong
authentication enables linking actions to user identities. Depending on
the level of accountability, these identities may be virtual identities,
such as usernames and/or links between multiple virtual identities, or
identities of the human users, systems, or applications running on
behalf of the users.

A seemingly conflicting requirement for accountability is privacy
protection. Users may not be willing to share their identities with
other users or computer systems. Similarly, user actions may not be
linked to real users. For example, in oppressive regimes, human lives
may be at risk if someone carries out an action that is against that
regime. The problem is further complicated in distributed and autonomous
systems without a trusted identity management capability.

In earlier works \citep{Farkas2002,Ziegler2006} it was shown that one can
achieve both privacy protection and accountability even in the absence of
an untrusted identity management server. A community-based identity
management and accountability framework was proposed. This system only allows the
users' virtual or real identities to be revealed if a quorum of the community
members agree to it. The basic assumption is that members of a community trust
the community to make the right decision. The method provides a proven
cryptographic approach to identity management. However, potential privacy
breaches due to data analytics and behavioural modelling of the community members
were not addressed.

\cite{Graf2023} aims to define accountable universal composability, focusing on
the modular analysis of the accountability framework. However, the privacy impact
of the rapidly evolving Artificial Intelligence (AI) systems has not been
sufficiently studied.

To address accountability in AI systems, information about
data quality and provenance must be available and trustworthy. This requirement is
difficult to achieve when dealing with data collected from open sources or 
untrusted (unverified) depositories. Moreover, federated computing and secure
multiparty computing may not support data provenance recording. During the last
decade, several data provenance approaches have been proposed to support a
variety of applications, from classical databases \citep{Buneman2019} to tracing
misinformation \citep{Allcott2017}, from scientific workflows \citep{Alam2022}
explainable AI \citep{Kale2023}. The provenance data must be protected from unauthorized and unethical modifications.
Technologies, such as blockchain \cite{NIST} are proven to provide reliable and
secure integrity protection of data and metadata.

\subsection{Ethical issues}
Research has shown the unintended negative consequences emerging from over-reliance on AI systems in sensitive decision-making contexts, without a clear understanding of the limitations and safe extent of the use of the various models \citep{nassar2021ethical}. In the criminal justice system, AI has been used for recidivism prediction.  AI helps to assess the likelihood of a convict re-offending, thereby informing parole decisions and rehabilitation efforts. Predictive policing uses AI algorithms to analyse historical data and predict potential future crimes.  This may help to allocate police resources effectively \citep{farayola2023ethics}. However, these systems have been producing disparate results and are prone to discriminating against vulnerable minorities.  For example, cases of scoring African-American defendants with a false positive score of higher likelihood to recommit a crime than white defendants have been documented \citep{washington2018argue, lagioia2023algorithmic}. Such disparate results have been caused by the problematic data quality selection for the algorithm, which scored based on the neighbourhood of provenance and family history in crime, which are both outside of the individual responsibility \citep{washington2018argue, lagioia2023algorithmic}.
Therefore, such systems could reproduce harmful biases when used in these sensitive decision-making contexts.

In the domain of human resources, AI assists in screening resumes, predicting candidate success, and even monitoring employee productivity, and has been deployed to attempt more data-driven hiring and management decisions \citep{kochling2020discriminated}. Nonetheless, such systems have been perpetuating biases in terms of gender, ethnicity, sexual orientation, or other characteristics when selecting candidates \citep{kim2016data}.

In the finance sector, AI-driven models have been deployed for credit scoring, fraud detection, and algorithmic trading, analyzing vast datasets to identify patterns and insights that humans might miss. These applications, however, come with significant ethical and fairness considerations, as the data and algorithms used can perpetuate biases and lead to unequal treatment of individuals \citep{hurlin2022fairness}.

Therefore, while AI can perhaps enhance decision-making processes, it is crucial to ensure transparency, accountability, and fairness in its deployment.

\section{Assistive Artificial Intelligence framework}\label{aai-framework}
In this section, we describe an Assistive AI framework that we envision. For the sake of simplicity, oftentimes we refer to this system as if it was already realized. It should be noted that this Assistive Artificial Intelligence framework is a high-level abstract concept. Some of the building blocks of this Assistive Artificial Intelligence framework are already present in state-of-the-art methods and tools that we partially overview in the Appendix~\ref{appendix}). Implementation (at least in part) can be the subject of future work.

\subsection{Preliminaries}\label{the-design-of-the-system}

\emph{Communication} is defined as 
information which is not private anymore, or, a \emph{product} becoming
available to customers. If the \emph{communication} is available to
everybody, then it is public. We propose a system of appropriate
procedures for (i) credible and accurate estimation of product
characteristics and quality, and (ii) rules limiting and controlling
product use, through legal redress for the maker of the product, i.e.,
accountability, \emph{without} infringing freedom of expression and the
protection of personal rights.

The principles are illustrated through examples. Modern computing and
AI methods complement these illustrations, and we
give examples in a separate section. They present the potential uses of
Assistive AI today and show some of the current shortcomings
highlighting future research and development challenges.

\subsection{Main components of the
system}\label{main-components-of-the-system}

\subsubsection{Knowledge base}\label{knowledge-base}

The Knowledge Base component holds the governing rules as well as facts of the community that utilizes Assistive Artificial Intelligence. Instances of this component are distinguished based on the type of knowledge they hold as well as the Board and its level of authority they belong to.

The two types of knowledge we would like to represent are factual knowledge and rules.

Factual information serves as a cornerstone in decision-making processes. This category is diverse, encompassing a wide range of facts that are instrumental in guiding judgements and conclusions. Specifically, these facts can be drawn from legal precedents, including both statutory provisions and individual case law, which provide a rich repository of legal reasoning and findings to inform future deliberations. Additionally, this category includes what might be termed as accepted or common-sense evidence. This spans historical facts, which offer context and background for understanding current events and decisions, and scientific statements, which are underpinned by empirical research and evidence, lending credibility and reliability to the conclusions drawn from them.

The selection of relevant or necessary facts is specific to the particular domain or community in question. Different fields or areas of study may prioritize certain types of facts over others based on their relevance, applicability, and the norms that govern that sphere of knowledge. For instance, in the legal domain, the emphasis might be on case law and statutory texts, while in scientific communities, the focus would be on data derived from rigorous experimentation and research.

Moreover, these facts are dynamic: they are influenced by new discoveries, changing social norms and changes in the legal framework. This calls for a flexible and adaptive approach to knowledge representation and inference so that the knowledge base system can adapt to changes and check the consistency of the knowledge space occupied by modified and new facts.

The term "rule" may carry different interpretations, which merit further exploration within the context of the law. In our framework, a rule is the description of the required social conduct or its
result. Rules shall be derived a)~directly from normative legal instruments
(written or unwritten constitutions, laws, [acts] and other regulatory
documents) produced by legislators or b)~from judgements of courts and other
decision-making instances (like different regulatory agencies). The latter could
be (i) judicial or official interpretation of legal norms and standards
\citep{maccormick1991interpreting} or (ii) a result of a decision related to
"hard cases" (i.e., cases "in which the result is not clearly dictated by statute
or precedent" \citep[1057]{dworkin1975a}. Decisions in hard cases very often are a result of weighing and setting up priorities between conflicting fundamental rights, e.g., freedom of speech versus human dignity.

\textbf{Legal Rules and their relations to Assistive Artificial Intelligence}  

When developing the knowledge base of Assistive AI, we conceptualize the law as a hierarchical system of behavioural rules \citep{kelsen1967a} in which the rules described in the legal system's normative documents (made specifically for regulatory purposes) are given concrete forms by the decisions of judicial forums and other bodies with decision-making powers. We distinguish three main levels of rules in the system: the level of fundamental rights (constitutional rights), rules of conduct, and regulatory norms (the distinction between rules and regulations, see above) contained in other legislation and individual decisions.

\emph{Fundamental rights} are at the top level in the hierarchy of rules. In most cases, written constitutions contain them. Their role is twofold: on the one hand, if the rules of the legal system come into conflict with each other, they help resolve the conflict. On the other hand, fundamental rights serve as a yardstick in the interpretation of generally formulated rules
\citep{dworkin1986a,dworkin1978a}. Ethical rules \citep{ai2019high} concerning AI are not included directly in our framework, but rules, especially rules of fundamental rights incorporate ethics in an indirect way.

\emph{Legal rules} and \emph{regulations} are closely interrelated part of the system. Since Assistive AI is a general concept, these rules can be diverse, depending on the particular domain the Assistive Artificial Intelligence is supporting. We have to distinguish between two types of normative structures, law and regulation. Although there is no sharp line between the two norm types, the law is more based on social evaluation and consideration of interests, while regulation more functions like a rule-based system (or an algorithm, in certain cases), which prescribes clear-cut procedures for social coordination, does not necessarily reflect values, and implements direct policy of the regulatory power.

The use cases of Section \ref{sample-use-cases} contain a mixture of legal and regulatory tools. The Assistive AI should be able to handle both normative structures to generate recommendations in a particular situation.

The judicial decisions and decisions of other conflict resolution instances (hereinafter both will be called judicial decisions) provide examples of rules, but sometimes they act as rules themselves, especially in case law (common-law) systems
\citep{holdsworth1934a,duxbury2008a,stone1959a,allen1958a}. In different legal systems, this rule-character of judicial decisions prevails to a different extent. Cases play an important role in the so-called civil law (continental) systems and in these systems, their interpretative (explanatory) character is stronger \citep{Hart1961,tiersma2006textualization}.

One may hope that a high-level  Assistive Artificial Intelligence system will be capable of suggesting the formulation of rules from a set of facts if it is desired. We outline this in Section~\ref{sec:knowledge_base_update}.

One may also differentiate between knowledge that can be modified by the Board of a community, the "Internal Knowledge Base" and knowledge that is external to the community forming the "External Knowledge Base". The "Policy of the Board" refers to the rules contained by the Internal Knowledge Base.  The External Knowledge Base includes "Law" and "General Facts" as those rules and facts are superior to the community and cannot be affected by the Board of that community. Such knowledge bases involve the external legal ecosystem of the community as well as common sense facts and judicial decisions that cannot be changed by the community.

\subsubsection{Author}\label{author}

The author is an entity that communicates (e.g., produces products), and may have a historical d-credit score based on previous communications and the related d-credit scores. The d-credit score, according to our definition, is domain dependent. Overlapping domains are possible. The entity that authored a communication can be:
\begin{itemize}
\item
  "fully anonymous" if nobody can provide evidence that a
  communication belongs to a given entity unless the entity voluntarily
  chooses to reveal its identity;
\item
  "anonymous, but accountable", if the identity of the entity can be
  uncovered according to the rules of the community; and
\item
  "known", if the identity of the communicating entity is available
  within a community or to everybody.
\end{itemize}

Note that anonymous author accountability involves proving that a given person belongs to a given code. Care must be taken in this regard, as this task is becoming increasingly difficult due to rapidly evolving faking technologies.

\subsubsection{Communication}\label{communication}

Communications are at the centre of the Assistive Artificial Intelligence model. In contrast to AI
regulations, our system is not about controlling AI, but using AI to
evaluate, filter, or possibly control communications so that individuals and their communities can
make informed decisions.
A communication is an item produced by one or more authors and made
available for one or more other authors and/or machines. Communication intends to have consequences or 
impact on other authors,  communities, or even the whole society in many contexts, such as social, economic, legal, ethical and cultural. 
Communication may appear in diverse forms, can be multimodal and maybe
a product (text, audio, video, image, logo, etc.) 

In our work, communication
has a restricted meaning in the sense that its existence shall be verifiable.
Verification is possible if communication becomes indefeasible (e.g., a
standalone static version), or if the court decides and makes it such a
version. Modification of the meaning of any communication is a new communication. This
distinction seems necessary due to the frequent changes in
communications in social media.

Tags derived from the law or the rules of a community characterize the
content of a communication. A set of \emph{reliability scores} can be
inferred for each tag of the communication. These reliability scores are
predicted values and may change over time when new evidence becomes
accepted. As a specific example, the field of science may take
advantage, beyond metrics like h-index, of the list of publications
retracted by journals.

\subsubsection{Derived value-related
information}\label{derived-value-related-information}

Information collected and processed from historical facts and other
resources (including the Knowledge Base) and used for the estimation of
values are called
derived information. Typical examples include
\begin{itemize}
\item the reliability score, 
\item the credibility score,
\item the network of communicated information, their cross-references and
the belonging values,
\item quality tags of communication.
\end{itemize}
The Assistive AI System may be able to discover hidden (contextual) variables and
explanatory variables that may influence court decisions.

\subsubsection{Board}\label{board}

We are considering communities that have Boards. The Board is the decision-making body of the
community to catalyze and regulate the working of the
community according to its prescribed role if there is any. The Board is
the human component in decision-making as regards the reliability and
compliance of communicated information. The Board could be anonymous
but must be accountable, and must have a credit score that is related to
the reliability of their decisions and internal regulations (focusing on e.g., withdrawing anonymity, advertising, giving salaries, and
prizes, organizing activities, and so on). Going back to the example of
retracted publications, such an event may influence both the reviewers
of the publication and the Editorial Board, too. Board decisions can be
overruled by higher-order Boards (see below) or by the court. The
Board's decisions can be assisted by the Assistive AI System's
recommendations.

The Board has the right to define, e.g., create, modify, eliminate, and
maintain internal regulations. These regulations include the conditions
by which actions can be executed in an automatic fashion using AI.

Lawsuits could be filed against the Board. For example, uncovering the
human entity might be against the law. If the Board loses a lawsuit, its
credit should be reduced.

The internal structure of Boards can differ from community to community
and is defined by the internal regulations. From the point of view of
our description, the rules of the Board are relevant and may
contain rules on how to implement a hierarchical structure within the Board
itself. In turn, we consider the concepts of "Constitution" and
"Board" such that the rules of the Board make the Constitution of the
sub-Boards and form a hierarchy.

\subsection{Operations that can be performed by the Assistive Artificial Intelligence}
\label{operations-performed-by-the-assistive-ai}

\subsubsection{Estimation of the reliability of
communication}\label{estimation-of-the-reliability-of-communication}

\begin{figure}[h!]
    \centering
 
    \includegraphics[width=0.95\textwidth]{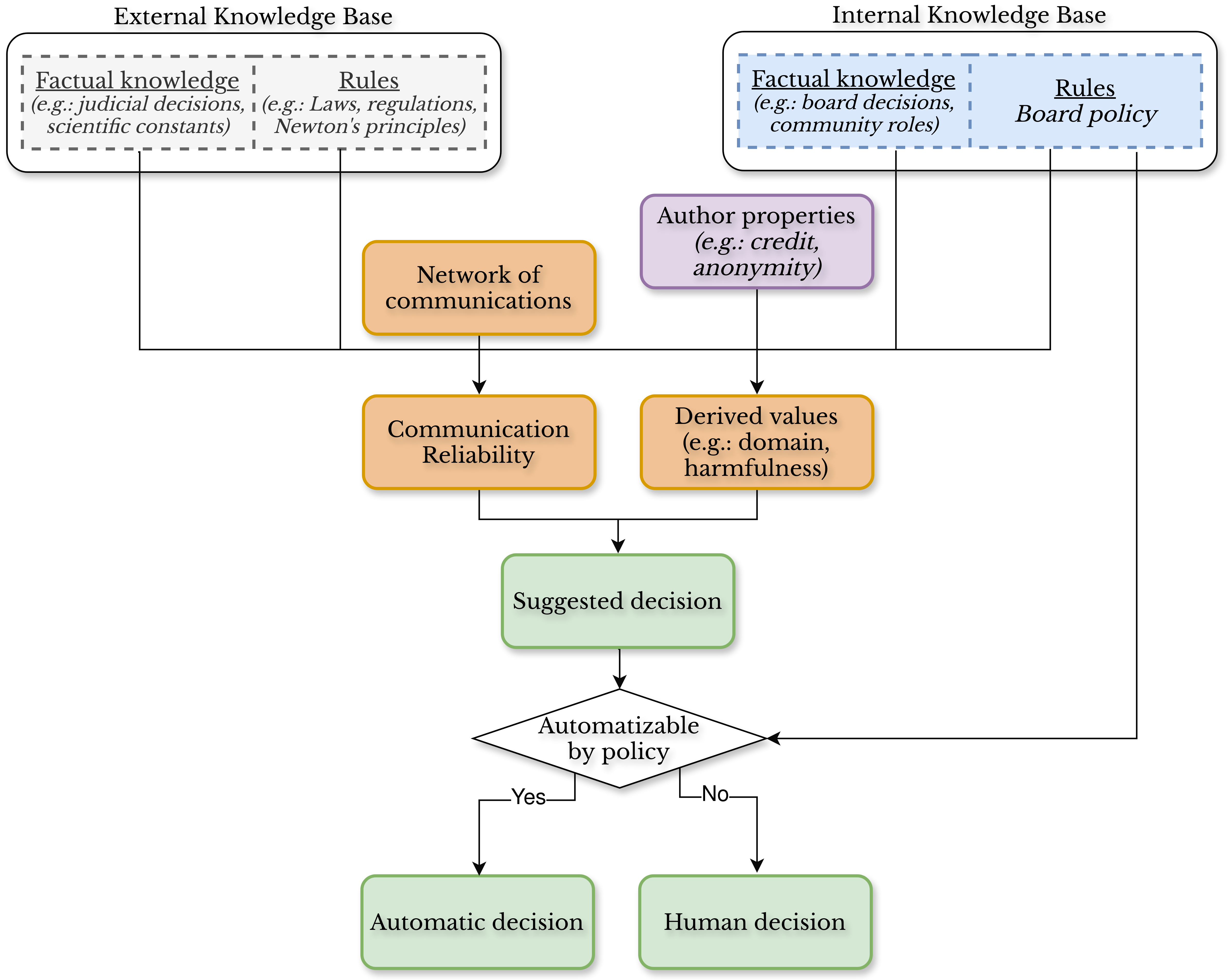}
    \caption{Flowchart for handling communications. The Assistive Artificial Intelligence system can suggest actions related to communications to the Board, to moderators designated by the Board and to end users. The final decisions can be made by human actors or, in certain cases, automatically. The choice of behaviour depends on the Board's policy.}
    \label{fig:communication}
\end{figure}

We consider the reliability of communication in the sense of its consistency with known information (such as external and internal Knowledge Bases, other communications, the reliability of authors, and common sense).
The reliability score of every communication is estimated. The method of estimation can depend on the community and the Board decides about it. In general, the method of estimating the reliability score may depend on different components. Some of them, such as the legal environment, for example, are necessary. Preferred components are listed below (see also Figure \ref{fig:communication}):
\begin{itemize}
\item
  the legal environment;
\item
  the Knowledge Bases;
\item
  cross-references and network information;
\item
  author's properties (anonymity, d-credit);
\item
  logical operations and commonsense;
\item
  contents of the communication.
\end{itemize}

\subsubsection{Qualification of communications}\label{qualification-of-communications}

Each communication is tagged with multiple category tags having
component values in $[0, 1]$ (i.e., continuous) or $\{0, 1\}$ (i.e.,
discrete). These tags are not to be displayed automatically, they serve
as intermediate variables used by the Assistive AI System and the
Board.

The set of category tags is based on legal categories, and seems advantageous to include other categories, such as "spam", "possible violation of privacy", and "contains personal data".

Qualifications include the elements of the list used for computing the reliability
score, plus, the reliability information. Tags can be used to \emph{draw
the attention of the user and the Board} as determined by the Board and
make \emph{recommendations for decisions} (with possible non-null
actions).

Example use-cases of quality tags are as follows:
\begin{itemize}
\item
  examples related to disclosing personal data;
\item
  description of a drug (which is a communication) and the qualification
  of the drug;
\item
  relevance of the communication concerning the given domain (such
  as drug effectiveness, or AI performance).
\end{itemize}

\subsubsection{Recommendation for
decision}\label{recommendation-for-decision}

Each recommendation contains a set of possible actions. Actions can be 
triggered, modified or eliminated by the human (user or the Board).

Recommendations and possible actions are described by:
\begin{itemize}
\item
  the legal framework (prioritized);
\item
  internal regulations (e.g., internal resolution of anonymity);
\item
  if they comply with the legal framework, including the precedent
  driven "ethical standards" that become common practice in legal
  decisions.
\end{itemize}

Assistive AI may bring attention to and make recommendations concerning
regulations of the community if they seem to conflict with the law due to the
context of the communication. For example, there are rights that cannot
be revoked or given up by internal regulations.

\subsubsection{Execution of automatic actions}\label{execute-automatic-actions}

Based on the evaluation and categorization of a communication, its reliability, and
the related regulations in a community, automatic actions can be made possible.

This is the case if no human intervention is needed (the task is
simple enough to be automated, or it is not significant, or the possible outcomes of the automated decision are checked thoroughly) a human decision here is not needed, or it is too slow, or both. The
extent to which decision-making is automated by algorithms is decided by
the Board. Otherwise, in case of conflict, a decision is recommended to
the involved parties as mentioned above. The Board, and possibly the
author of the Communication, or other affected users are notified about
the automatic decisions when they are made. An example is if the communication is tagged as "hate speech". Then the communication is deleted and both the author and the Board are notified. In case of an indication that the communication may be "hate speech", e.g., by common sense reasoning, the author and the Board are notified.

\subsubsection{Monitoring the knowledge base and informing the Board of any
relevant changes, discrepancies, and discovered
dependencies}
\label{sec:knowledge_base_update}

Legal rules are constantly changing, so updating mechanisms must be built into the system. We detail three important update mechanisms without aiming for completeness

\begin{enumerate}
    \item One of the update mechanisms monitors changes in legislation and may continuously incorporate new and changed rules into the knowledge base.  
    \item Another mechanism monitors the decisions and, if new rules or new interpretations of rules incorporated in legal rules and regulations appear in the decisions, it is also incorporated into the knowledge base. If a decision is significant it becomes a precedent that can be referred to in later decision making. Step by step, such decisions may become the practice, and finally part of the legal system. Since the judicial practice is the first to face life's problems, most of the time judicial decisions provide the first solution to these problems. In common law systems, these decisions will be the so-called landmark cases. In continental legal systems, some decisions are also included in the text of legal acts. 
    \item At a lower level, a community can have a decision-making body, i.e., its own Board, and the Knowledge Base can be extended by the internal regulations or decisions and recommendations of that Board.
\end{enumerate}

The Assistive AI system may -- and actually should -- use certain legal rules as enshrined in AI regulations as starting points for helping timely and well-informed decision-making.  The focus of the system is on allowing communities to manage AI through their Boards so that it can help them in individual or common decision-making processes in line with their own goals, conditions, and considerations among other things.     

The Assistive AI System monitors the Knowledge Base. If it finds any rule candidate that could be derived from the Knowledge Base it proposes that candidate rule to the Board. The Board decides on the validity and desirability of the suggested rule, and then
it either accepts it or rejects it. If the rule is accepted, then it is added to the Knowledge Base. On the other hand, if the rule is rejected, then the data supporting the AI's inferencing may be modified. For example, the relevance or reliability of the data may be changed in such a way that the AI will not recommend this rule further.

\begin{figure}
    \centering
    \includegraphics[width=0.85\textwidth]{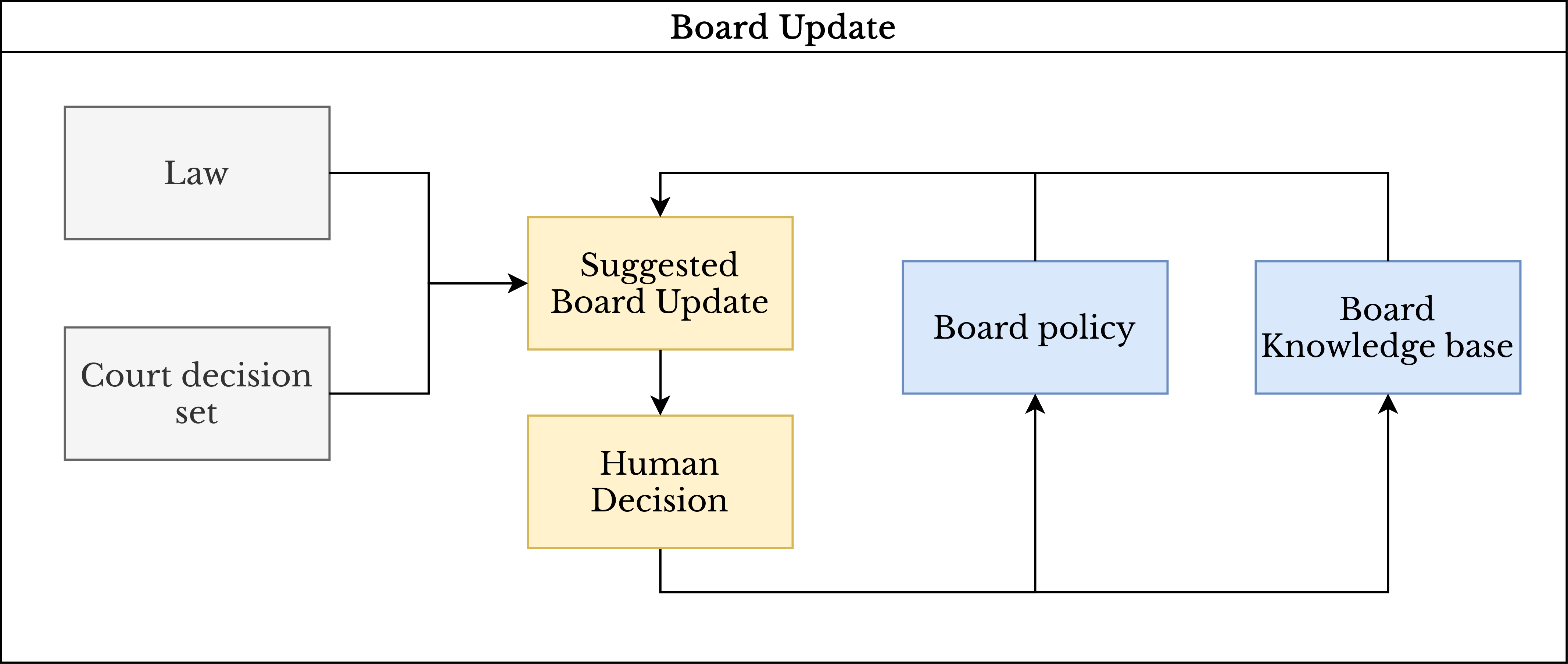}
    \caption{Flowchart for updating the Board knowledge base and policy based on Assistive Artificial Intelligence recommendations. Box colours in this figure indicate component levels, legal (grey), board (blue), and decision level (yellow).}
    \label{fig:board_knowledge_base}
\end{figure}

Modification of AI models and processes might be needed in both cases.

If the Assistive Artificial Intelligence system is well defined and it detects a likely conflict between the internal rules and
the legal framework then it should report the discrepancy to the Board or the users, including the authors and
the readers for resolution. This way the Assistive Artificial Intelligence system becomes a legal assistant to the users and their communities.

\subsubsection{Keeping derived variables, and values up to
date}\label{keep-derived-variables-and-values-up-to-date}

The list of derived variables might be extended. The Assistive AI System
should be able to keep these variables up to date. Reliability score and
d-credit calculation, or explanatory variable search is a sub-process of
this general one with extra significance and added details. In our
current case, this general process should include the updating
procedures of the network-like database of communications and their
cross-references as well (see Figure \ref{fig:board_knowledge_base}).

\paragraph*{Domain-dependent credit
calculation}\label{domain-dependent-credit-calculation}
\addcontentsline{toc}{paragraph}{Domain-dependent credit calculation}

\begin{figure}
    \centering    
    
    \includegraphics[width=0.95\textwidth]{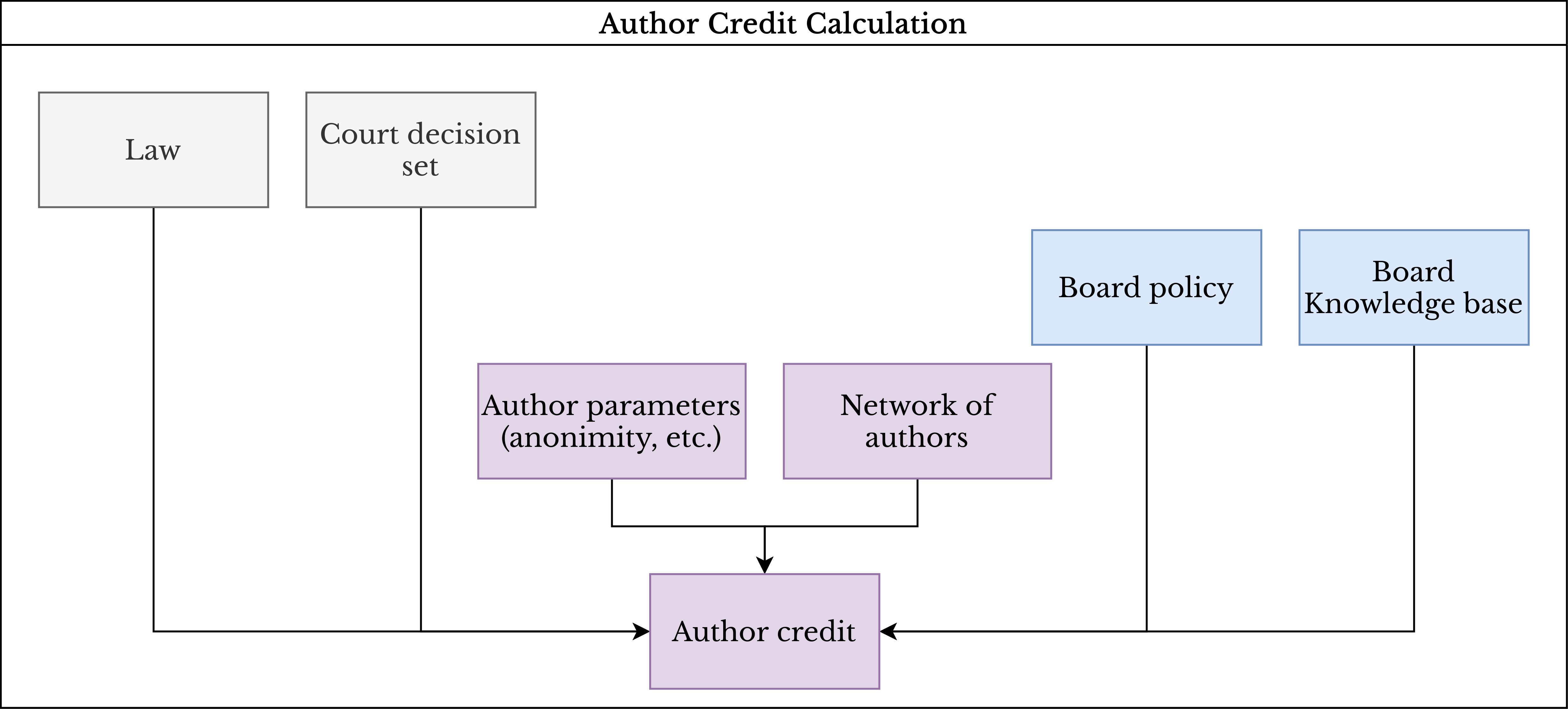}
    \caption{Author credit calculation. The calculation method takes into consideration the Board's policy, knowledge base, and relevant legal requirements. Concrete credit scores, which can be domain-specific, are computed based on individual author properties (anonymity, authored communications etc.)\ and the author network.}
    \label{fig:author_credit}
\end{figure}

The calculation methodology of the author's credit score (see Figure \ref{fig:author_credit}) depends on
internal regulations. We believe that it should be domain-dependent.

The domains may be described by a taxonomy
tree with values on the leaves and any communication may belong to several
leaves. Other methods may also be introduced by the Board. 

\section{Sample use cases}\label{sample-use-cases}

In the following sections, we present certain instances of the proposed Assistive Artificial Intelligence system in different domains, such as recommendation on advertisements of drugs and health services, scientific communication, and financial product recommendations. A high-level overview of related AI research and technology components that could be utilized in practical realizations is presented in the Appendix (Section \ref{appendix}).

\subsection{\texorpdfstring{Recommendation on advertisements of drugs or
health services
}{Recommendation on advertisements of drugs or health services }}\label{recommendation-on-advertisements-of-drugs-or-health-services}

The model of Assistive Artificial Intelligence can be used in areas of commercial communication where providing the public with authentic and reliable information affects particularly important individual and social interests. The area of advertising drugs, medicinal products and health services is clearly such a domain. The importance of this domain is increased not only by the value involved, i.e.\ human health but also by the fact that the domain represents an ever-larger slice of commercial communication. Another relevant factor is that in recent years, especially in connection with the COVID-19 pandemic, this domain has become a crucial arena for information battles and disinformation. The use of Assistive Artificial Intelligence in this domain is supported by the fact that commercial communication in general, and pharmaceutical advertisements in particular, are subject to more legal regulations and recommendations than other areas of social discourse, thus the criteria for evaluating communication are clearer.

The purpose of using Assistive Artificial Intelligence in the domain of drug advertising is to give the customer individualized recommendations/opinions concerning the reliability/correctness of the advertisement of a particular drug, medicinal product, or health service. In a broad interpretation of advertising, the domain covers all communication that addresses the public with information about a particular drug, medicinal product, or health service. The author of communication can be in the domain, above all, the manufacturer or distributor of the given product or service, or the person who publishes an advertisement or shares substantive information about it.

Assistive Artificial Intelligence assesses the reliability cq.\ authenticity of the information for the recipients of the communication, i.e.\ for consumers. In this context, it evaluates the credibility of the communication on the one hand, based on the data and information regarding the content of the communication and the advertised product. On the other hand, it also evaluates the reliability of the informant (advertiser) on the basis of relevant historical data, for example referring to previous legal disputes or credibility problems. The evaluation carried out by Assistive Artificial Intelligence can be individualized: by including a given consumer’s data, the system can make a recommendation that takes into account the individual consumer's aspects.

As mentioned, the knowledge base in the domain can be richer compared to other areas of social discourse: from the regulatory side, regulations, recommendations and good practices on the advertising of medicines can all be part of it.

Multiple boards can be envisioned within the domain that can assume responsibility for the operation and supervision of the Assistive Artificial Intelligence. Consumer protection organizations can be as interested in operating such an assistive system as organizations that support the enforcement of patient rights or represent particularly affected social groups (e.g.\ the elderly).

Current state-of-the-art offers multimodal foundation models for processing mixed modality media, such as commercial communications with graphical and textual content. Table~\ref{tab:academy} identifies the main elements of this domain as elements of the proposed Assistive AI framework.

\rowcolors{2}{gray!25}{white}
\begin{table}[h]
\centering
\caption{Domain-specific elements as abstract building blocks of the Assistive AI framework for recommendations on advertisements of drugs or health services.}\label{tab:adv_drug_health}
\begin{tabular}{
  >{\raggedright\arraybackslash}p{0.24\linewidth}
  >{\raggedright\arraybackslash}p{0.24\linewidth}
  >{\raggedright\arraybackslash}p{0.48\linewidth}
}
\toprule
Elements & Architectural Role & Description, properties \\
\midrule
Advertiser & Author & The advertiser of the particular drug or medical service (pharma companies, health service providers, etc.) \\
Consumer & Recipient of communication & Individual who is receiving the commercial communication \\
Advertisement & Communication & An advertisement of a drug or health service in any media \\
Consumer protection association / patients' rights association & Board & Regulations, recommendations concerning drugs and health services \\
Upper Body & Regulation & Laws concerning drugs and health services \\
\bottomrule
\end{tabular}
\end{table}

\subsection{Opinion about the credibility of communication on academic
subjects}\label{opinion-about-the-credibility-of-communication-on-academic-subjects}

A possible use of the Assistive Artificial Intelligence model is for evaluating a communication or a particular person forming an opinion on academic subjects or matters of public interest in public spaces, vindicating academic authority. Different situations can be imagined. During the COVID epidemic, for example, the number of experts or pseudo-experts speaking in public spaces about various subjects related to the pandemic, (effectiveness of vaccination, spread and risk of the epidemic etc.)\ increased enormously, and in the huge noise, it was very difficult to decide which opinion and which expert is the one whose opinion is reliable.

Another example is the discussions and experts related to artificial intelligence. This area is even more problematic compared to the communication related to COVID, as it is more difficult to identify the profession that can serve as a reference point when judging a communication. Both the press and public spaces are full of experts who give completely divergent opinions and predictions about the dangers, uses, and social effects of artificial intelligence. Given the complex nature and rapid evolution of the field, some opinions may identify potential opportunities and risks, for example, after thorough research on unemployment expectations in different occupations, while others may be baseless guesses only.

The goal of the Assistive Artificial Intelligence use in this domain is to give an opinion (recommendation) about the academic credibility or authority of public speakers, and particular pieces of communication. 

Assistive Artificial Intelligence collects information on the content of the communication and its subject (including academic views and guidelines) on the one hand, and about the speaker -- about their academic track record, previous public appearances, and their evaluation, problems, and contradictions, if any, and about the relevant legal and regulatory environment (where appropriate). 

The Board in the context of this use case could be any academic community (research groups, institutes, departments, universities, associations) that has a proven track record in an academic field. Due to the quick development and the diverse views even in the academic community, tracking the fate of publication there is a need for the fast updating of d-credits of experts and Boards. RetractoBot, the research project of the Bennett Institute for Applied Data Science at the University of Oxford is an intriguing example of such effort \citep{retractobot2023}. This method, which automatically emails authors, when papers they have cited are retracted, could serve as a basis for the Assistive Artificial Intelligence d-credit update method. Table~\ref{tab:academy} identifies the main elements of this domain as elements of the proposed Assistive AI framework.

\rowcolors{2}{gray!25}{white}
\begin{table}[h]
\centering
\caption{Domain-specific elements as abstract building blocks of the Assistive AI framework for assessing the credibility of communication on academic subjects.}\label{tab:academy}
\begin{tabular}{
  >{\raggedright\arraybackslash}p{0.24\linewidth}
  >{\raggedright\arraybackslash}p{0.24\linewidth}
  >{\raggedright\arraybackslash}p{0.48\linewidth}
}
\toprule
Elements & Architectural Role & Description, properties \\
\midrule
Public speaker & Author & Person forming opinion about academic subjects or matters of public interest in public spaces vindicating academic authority. \\
Listener & Recipient of communication & Audience of the public discussion. \\
Opinion & Communication & Opinion vindicating academic authority. \\
Academic community & Board & Any size of academic communities (research groups, institutes, departments, universities, associations etc.) with internal guidelines, and rules of the scientific community. \\
Upper Body & Regulation & Laws, regulations, scientific guidelines, and requirements.\\
\bottomrule
\end{tabular}
\end{table}

\subsection{Financial product
recommendations}\label{financial-product-recommendations}

Financial Product Recommendation Systems (FPRS) with Assistive Artificial Intelligence are nowadays increasingly important in the financial sector on the reason they help financial institutions provide personalized and relevant recommendations to their consumers. However, there are challenges and limitations, e.g.\ data privacy and security concerns, biases and ethical issues, and cold-start problems and scalability issues even though they could help financial institutions provide personalized investment recommendations, automate credit decisions, and offer personalized insurance policies.

When the data used to train these systems fails to represent the entire user population, biases might arise, resulting in unfair and discriminatory recommendations. For example, an FPRS that recommends credit cards only to consumers with high credit scores may exclude those who have a low score but may still benefit from using the card. FPRS should follow ethical guidelines and not recommend products or services that are detrimental to the consumer's financial well-being.

Implementing and operating an FPRS shall be a complex process, including data collection, algorithm selection, and user interface design. Upon algorithm selection, the algorithm shall be implemented and integrated into the financial institution's existing infrastructure. The whole system should be tested and evaluated to ensure that it is accurate and effective in providing personalized recommendations to consumers. Monitoring and evaluation should be ongoing since they are important to ensure that the FPRS remains effective as user behaviour and preferences change over time. The board is to address and oversee monitoring, evaluation and taking action to ensure ethical guidelines and lawfulness. Table~\ref{tab:finprod} identifies the main elements of this domain as elements of the proposed Assistive AI framework.

\rowcolors{2}{gray!25}{white}
\begin{table}[h]
\centering
\caption{Domain-specific elements as abstract building blocks of the Assistive AI framework for financial product recommendation.}\label{tab:finprod}
\begin{tabular}{
  >{\raggedright\arraybackslash}p{0.24\linewidth}
  >{\raggedright\arraybackslash}p{0.24\linewidth}
  >{\raggedright\arraybackslash}p{0.48\linewidth}
}
\toprule
Elements & Architectural Role & Description, properties \\
\midrule
Customer & Author & Income, age, number of dependants, wealth, guarantor, \ldots{} \\
Loan application & Communication & Parameters of the desired loan \\
Financial Institution & Author & Related financial products, financial results (publicly available), FSA's data on the institution \\
Financial product & Communication & APR, loan amount, term, target group, interest period \\
Ethical Banking Guidelines & Board & Extra norms, rules, guidelines. Requires accountability for financial institutions \\
Financial Supervisory Authorities (FSA) & Legal entity & Norms, rules, guidelines, standard processes, decisions \\
\bottomrule
\end{tabular}
\end{table}

\subsection{Financial credit scoring}

AI in credit scoring involves using machine learning algorithms to assess creditworthiness more accurately than traditional models.
Business description and intent:
Banks focus their efforts on harnessing artificial intelligence to refine its process for making credit decisions. According to the McKinsey Global Institute, the potential financial impact of machine learning solutions for risk assessment in the banking industry exceeds USD\,250 billion. 
Use case details:
Assistive AI increases both the accuracy and the inclusiveness of credit assessments, especially aiding customers with sparse or non-traditional credit histories. The Assistive Artificial Intelligence-supported technology analyses not only traditional credit history but also non-conventional data such as spending behaviours as an alternative data source. With a Board specialized for Assistive Artificial Intelligence human control in the scoring process, we assure quality:
\begin{itemize}
\item	leveraged diverse data such as work experience, income, transactions, and credit history to generate personalized and reliable credit scores;
\item	eliminated emotional bias for equitable evaluations, removing gender, racial, and other biases, resulting in faster and process-controlled loan approvals and enhanced customer convenience.
\end{itemize}
Potential results:
Assistive Artificial Intelligence has led to the conditions of more customized credit solutions, a reduction in default risks, and the extension of credit access to a wider range of customers. Machine-learning models can handle large and complex data sets and can detect patterns and trends that may not be visible to human analysts without AI. In this way, banks will know more, customer transparency will grow and the level of transparency will need scrutiny. 

This innovation, the integration of a Board in credit scoring practices will position human control as a more inclusive and progressive role in the course of automated decision preparation in the financial services industry. Human control via a Board ensures that the best practices and principles of credit scoring will be followed by keeping laws or norms. Moreover, it provides guidelines to improve performance quality to balance the benefits and harms of automation to incorporate the value of human judgment in credit scoring. Table~\ref{tab:fin_credit} identifies the main elements of this domain as elements of the proposed Assistive AI framework.

\rowcolors{2}{gray!25}{white}
\begin{table}[h]
\begin{threeparttable}
\centering
\caption{Domain-specific elements as abstract building blocks of the Assistive AI framework for financial credit scoring.}\label{tab:fin_credit}
\begin{tabular}{
  >{\raggedright\arraybackslash}p{0.24\linewidth}
  >{\raggedright\arraybackslash}p{0.24\linewidth}
  >{\raggedright\arraybackslash}p{0.48\linewidth}
}
\toprule
Elements & Architectural Role & Description, properties \\
\midrule
Scoring Agency & Board / Author1 & Describes and oversees the method of financial credit scoring. (This process can’t be fully automated, see, e.g., the SCHUFA case\tnote{1}.) \\
Financial credit score & Derived value of communications 1 and 2 & Numerical value to be forwarded to the financial institution (external to the "scoring" community we model here). \\
Loan applicant & Author2 (see \ref{financial-product-recommendations}) & Person who applies for a loan. \\
Loan application & Communication2 (see \ref{financial-product-recommendations}) & Parameters and history of the applicant.\\
Background research about the applicant by the scoring agency & Communication1 & Additional information about the applicant (should be included in scoring).\\
\bottomrule
\end{tabular}
\begin{tablenotes}
  \small
  \item[1]Judgement of 7 December 2023, OQ v Land Hessen ("SCHUFA Holding I"), Case C-634/21, EU:C:2023:957. \url{https://curia.europa.eu/juris/documents.jsf?num=C-634/21&language=en} accessed on March 13, 2024.
\end{tablenotes}
\end{threeparttable}
\end{table}

\section{Discussion and future work}\label{discussion-and-future-work}

In 2003 already, the U.S. National Science Foundation issued a call concerning potential societal transformations that can be triggered by emerging IoT technologies. It expressed concerns about the future \citep{nsf-03-611}.  The call highlighted that "information technologies are designed, used and have consequences in many social, economic, legal, ethical and cultural contexts". These apprehensions have been significantly amplified with the progression of artificial intelligence. 

Today, the pervasive presence of AI in everyday life has already led to frequent misuse, posing substantial societal risks and challenges. In response to these concerns, regulatory measures are being initiated. Notably, the European Union has taken a pioneering step by issuing the first regulation on artificial intelligence on March 11, 2024 \citep{Madiega2023}. While such regulatory efforts are crucial, they may not fully address the problem due to the following limitations: (a) although legal disputes over rule violations can be taken to court, the judicial process tends to be slow, and (b) the rapid pace of AI development means that regulations, which may already be imperfect and require adjustments, cannot keep up. Consequently, significant and potentially irreversible damage may occur in the interim.

We put forth the concept of Assistive AI integrated within a legal framework as a proactive credibility estimating instrument for supporting lawful, trustworthy, and informed human decision-making. To this end, we propose a flexible, domain- and community-dependent AI framework made of an extendable and partially organized network of Boards. The term "lawful" implies that the regulations set by any Board must align with the constraints of the rules of higher-order Boards and, consequently, the Constitution itself. Building trust is facilitated through the rapid AI-assisted assessment of communication reliability, which serves as a complement to the inherently slower mechanism of establishing accountability dictated by regulatory acts.

However, consider scenarios where legal violations are noticed but contractual obligations restrict the revelation of such information. This dilemma can be resolved by anonymous yet accountable communication, where anonymity protects the communicator unless the disclosed information proves to be false, necessitating accountability. The reliability estimation depends on accountability: (a) if the author of the communication is not accountable then the reliability of the information may be considered low or even nullified, as per the rules of the relevant Board and (b) otherwise, reliability can be estimated based on the author's domain dependent historical reliability referred to as d-credit here, alongside the trustworthiness of their network.

In our framework in every scenario, human oversight governs decision-making. In instances where the timeline or volume of decision-making needs exceed human capabilities, requiring rapid responses or managing numerous decisions, respectively, Boards may permit AI-driven decisions under rigorous conditions, in compliance with regulatory standards such as the AI Act of Europe. In all other situations, the decision-making process remains entirely in human hands.

We listed a few sample use cases (Sect.~\ref{sample-use-cases}) and included algorithms that could be applied to them. However, the deployment of assistive AI hinges on robust and validated algorithms. Section~\ref{ai-testing-safety-and-monitoring} delves into this topic, underscoring that while testing -- a cornerstone of traditional software development -- has achieved a level of maturity, comprehensive testing remains elusive as outlined by \citet{kaner1997impossibility}, except in rare instances with certain software languages like Ada, as discussed by \citet{knight2009echo}. For most software, we are limited to estimating reliability, a topic explored by \citet{zarzour2021sequential} and referred works. 

The case is more complex for AI architectures, particularly deep-learning models, where the intricacies of the process are obscured by the zillions of weights within artificial neural networks and the process is not transparent. Additionally, generative models introduce the risk of generating fictitious data, called "hallucinations". This necessitates a cautious approach to applying AI methodologies based on careful and tedious estimation of reliability.

Consequently, while the technological components exist, the actual construction of the architecture will require time due to the need for further advancements. These developments include establishing probabilistic reliability estimates for each AI component and making algorithmic enhancements when these targets are not met. It's possible to develop the architecture incrementally, though specific challenges may arise. For instance, methods for anonymous but accountable communication have been around for some time \citep{Ziegler2006}. 

In our Assistive Artificial Intelligence framework, we introduce key principles: (a) human decision-making with clearly defined exceptions, (b) the conditional anonymity of Board members based on specific Board regulations, and (c) the imperative for Board accountability, necessitating the verification of Board members as humans. These principles underscore the critical need for continuous progress towards Zero Trust Authentication \citep{stafford2020zero}, as a measure to safeguard against the increasingly sophisticated techniques employed by hackers.

The estimation of domain-dependent reliability, or d-credit, is a key component of the Assistive AI system serving fast searches for novel information and helping information selection. Depending on the goals and the rules of the Board, this estimation may be modulated by the network of related communications, e.g., the referring content of citing papers in scientific journals, or court decisions that enforce corrections, to name a few. Historical information about the d-credit of the communications of an author and the actions taken by the Board can be used to associate d-credit values to the author as well as to the Board. 

We acknowledge that the investment of the necessary resources for the development and maintenance of the Assistive AI system can be high. We posit that the potential damage to society resulting from malicious AI attacks that may arise due to the delays in countermeasures based on the regulatory processes could far exceed the operational costs of the Assistive Artificial Intelligence system.

\section{Conclusions}\label{conclusion}

In this paper, we sought the answer to the crucial question of how we can get reliable information to make informed decisions that are in line with community goals and values and are congruent with the law in an increasingly turbulent, uncontrolled information environment. Our proposed system of Assistive AI invites us to look at AI not only from the otherwise justified regulatory approach that sees certain risks and dangers in its spread but also as an effective tool that can help us make decisions on a sound basis. The elaboration of a workable model based on this recognition is of central importance since we already are and will increasingly be in need of using AI to tackle the vast amount of information and data that surrounds us. We argued that the use of AI to assist human decision-making is a practical and realistic idea if an appropriate framework can be designed. The goal of our proposed Assistive AI system is to meet the requirements of such a framework.

We contended that AI can help us in the evaluation of information and communication if we design its operation for sufficiently specific goals and aspects. The system delineated in this paper would ensure this through two elements. First, communication domains should (and can) be identified so that the operation of an Assistive AI can be adapted to the special characteristics and motivations of the domain. An Assistive AI can be used in any social communication framework (domain) where the criteria of communication (contained by the knowledge base) can be sufficiently specified. In our sample use cases the communication field of drug advertisements, financial product recommendations, or speech on academic subject could serve as examples for such domains.

Second, the focus of our proposed Assistive AI system is on social groups and communities, which -- within the framework of a given domain -- use AI according to their own interests and aspects. Assistive AI can be used for individual purposes and it is realistic to assume that smaller or larger communities will maintain and operate such systems to better guide their members (and those who share their goals and trust their system) in the abundance of information. If an adequate number of communities or social groups use the model and develop an Assistive AI that fits their criteria, then individuals will also be in a good position to rely on Assistive AI systems that best suit their personal goals and value choices in their decisions.

Closely related to all of this, we argued that AI can only serve as our helpful support in cleaning up communication if the enforcement of specific community interests and aspects in a given domain is ultimately ensured by human supervision. Accordingly, in our proposed system of Assistive AI, it is the human Board that has the final say on all relevant issues that define the system. Only AI can evaluate all acts of mass communication, but the adequacy of its evaluation work and the arising new systemic issues need to be judged by humans (the Board). An Assistive AI enforces the requirements of accountability on both sides: while the AI is responsible to the Board, the Board is responsible to the given community for the proper operation of the system according to legal requirements and community interests.

We believe that the proposed system of Assistive AI is a workable framework for exploiting the positive application opportunities of AI to those individuals and their communities who are looking for reliable guides in the increasingly complex digital reality. According to the idea of Assistive AI, whether AI can really do this job for us depends on whether there will be a sufficient number of groups (and individuals) who not only recognize the communication dangers lurking around them but are also willing to take action to eliminate them. Our idea of Assistive AI -- that is based on social activity -- can therefore only be realistic and effective if our communities have adequate resources to set up their own systems, and if they take advantage of this opportunity in sufficient numbers. With or without AI, there is no solution that could spare our own efforts.

\bibliographystyle{plainnat}
\bibliography{aai}

\appendix
\section{Appendix}\label{appendix}

In this Appendix, we showcase candidate models, methods, and tools from the rich arsenal of state-of-the-art literature that Assistive Artificial Intelligence could be built upon. They are listed in the context of the use cases of the main text. 

We start each use case with brief motivations about the methods that we include in that subsection. It is then followed by an overview of the listed methods. For the details of the methods themselves, the interested reader is referred to the literature.

\subsection{Recommendation on advertisements of drugs or health services}\label{recommendation-on-advertisements-of-drugs-or-health-services-ai-background}
Advertisements are complex, multimodal communications. They might contain audio-visual data and different types of textual information (such as statements, or tables). To assess the credibility of the advertisements the AI model must be able to extract specific entities (e.g., kind of drugs, laboratories) and statements (e.g., quality measures) from this multimodal data in a way that it will be robust and unaffected by deception.

\textbf{Multimodal foundation
models}

Recent advancements in LLMs provide a boost for the integration and
understanding of other modalities as well. These advancements highlight semantic
understanding capabilities of images, videos, audio, spatial information, and
motion \citep{Team2023,Girdhar2023,Jiang2023,Zhang2023}. The generalization
capabilities of such models are strengthened by language representation through
joint embeddings \citep{Yu2022}. Emerging modality alignments could also be
observed if a connecting modality exists during training \citep{Girdhar2023}.

Many popular solutions from state-of-the-art literature adopt the autoencoder
plus latent generative model architecture. This usually consists of two modules.
First, a vector-quantized variational autoencoder (VQ-VAE), that is trained to
encode the input to a dictionary-bound discretized latent representation, then
decode it with the smallest deviation possible \citep{Oord2018}. The role of
this VQ-VAE is to serve as an interface between the modality (such as images,
videos, audio signals) and the Transformer-readable sequence of discrete tokens.
The second element of such an architecture is the generative model that works on
the latent sequence (usually in the form of an autoregressive Transformer model
or a Diffusion model) \citep{Ramesh2022}.

These latent multimodal sequences could be placed as prefixes for the language
tokens or interleaved with the language tokens \citep{Liu2023,Team2023}.
Generating and decoding tokens from different modalities, as well as performing
conversational and instruction-alignment for these sequences is possible. This
capability could be utilized for generating common sense explanations and
refined answers in a conversational way analyzing multiple modalities
\citep{Zhang2023}.

Such applications are approaching the human level of multimodal understanding in
an automated way \citep{Team2023}.

\textbf{Relationship extraction}

Relation extraction (RE) is an important field of language and media processing.
To process multimedia content of advertisements elements of the communication,
such as the subject medicine, side effects, and pharmaceutical metrics should be
identified and properly linked. Relation extraction is the field related to
extracting attributes, entities, and their connections classically from text.
Latest advancements include multimodal relation extraction and LLMs that are
utilized and tuned for RE tasks \citep{Lin2023,Zheng2021,Sun2023a,Chen2023b}.
Demonstrations of state-of-the-art multimodal relation extraction (MRE) in the
biomedical domain have also been carried out lately, which implies that analysing
advertisements in the biomedical domain could be possible with a minor adaptation
of existing tools \citep{Zhou2023,Lin2023,Peng2023}. By performing MRE exact
statements could be extracted from commercial communication that usually targets
human audiences in natural language, audio, and video. Computer-based processing
and compliance analysis of the extracted relations make it possible to achieve
explainable and correct reasoning over the content and the rules set by the
Board or the Legislator.

\textbf{Deception of large language
models}

It is a well-known characteristic of LLMs that they can generate misleading
content that is often difficult to detect because of the models' ability to
seamlessly blend facts and fiction, and stylistic flexibility. Cases of \emph{LLM
hallucination}, i.e., LLM-generated deceptive output that is unintended by users
and operators, range from providing factually incorrect but contextually
relevant information to responses that diverge from the provided context, e.g.,
user instructions (factuality and faithfulness hallucinations, see
\cite{huang_survey_2023} for a recent survey). While LLM hallucinations are
unintended, the models can also be used to generate deceptive content
intentionally. LLMs can be provided with misinformation during fine-tuning or as
a generation context, which can lead to misleading output, and they can also be
explicitly trained or instructed to generate content to induce false beliefs in
the audience \citep{park2023ai,hagendorff2024deception}. Experiments show that
state-of-the-art LLMs like GPT-4 have emergent abilities to understand and
reason about false beliefs, and follow strategies to elicit them
\citep{hagendorff2024deception}.

\textbf{AI models for advertisement credibility
assessment}

Automatic detection of health and drug-related misinformation including
misleading or factually incorrect advertisements have become an intensively
researched topic in recent years, especially since the outbreak of the COVID-19
pandemic, which led to a large increase in the spread of medical information,
including misinformation (see \cite{schlicht_automatic_2023} for a recent survey). The
problem is typically approached as a text classification task, and solved either
by traditional supervised machine learning methods such as random forests or by
-- most frequently pre trained transformer based -- deep learning techniques,
which commonly achieve better performance \citep{schlicht_automatic_2023}. The used
methods require extensive training and test datasets, and the shortage of high-quality,
expert-annotated resources, especially for domains that are not COVID-related
\citep{ni_rapid_2023,schlicht_automatic_2023} has led to strong interest in efficient
dataset creation methodologies such as active learning \citep{nabozny_active_2021}.

\subsection{Opinion about the credibility of communication on academic subjects}
\label{opinion-about-the-credibility-of-communication-on-academic-subjects-1}

Academic authors and communications form a network. One layer is formed by authors. The other layer has the communications, i.e., the papers. Authors are linked to papers and papers refer to other authors through the papers they cite. To estimate the credibility of an academic author the AI system has to utilize this network-based information and propagate credibility between related authors while also taking their domain-specific scientific metrics into account. The AI should analyse scientific statements in communications via logical reasoning to identify their meaning as well as the context in which a given publication cites another work (e.g., how it builds upon it, or if it contradicts it, among other things). 

\textbf{Reasoning over logical
representations}

Reasoning over the content of scientific communications and assessing their internal consistency and logical relationships can be implemented using the tools described in detail by  section "Logical rule representation and inference" of Appendix  \ref{appendix:financial-use-case-ai-background}: Documents can be parsed into formal meaning representations with well-defined model-theoretic semantics (e.g. Uniform Meaning Representations discussed above), and these representations can, in turn, be analysed and reasoned over using suitable theorem provers and model checkers.

\subsubsection{AI applications in academic credibility and scientific
metrics}\label{ai-applications-in-academic-credibility-and-scientific-metrics}

Assessment of the credibility of academic communications, e.g., research papers
and technical reports, plays a vital role in science, both in discoveries and
the dissemination of knowledge. Important examples are peer reviews conducted by
academic journals and conferences, and the evaluation of research output by
government agencies distributing funding. Although final evaluations are made by
human experts, creating AI algorithms for \emph{assisting} assessment is an important research topic, and several tools are available for
automatically evaluating various aspects of scientific communications that
contribute to their overall credibility \citep{kousha_artificial_2024}. Solutions
were developed for finding plagiarism and checking whether references match
their citations (see \citet{qazan2023performance} for a survey), and there are
tools for checking the rigour, reproducibility and transparency of the described
research, including algorithms for estimating the plausibility of reported
statistical results \citep{nuijten__2020}. While solutions for this
type of credibility assessment support frequently use traditional feature
engineering and machine learning methods such as bag of word representations,
keyword extraction and linear models or decision tree ensembles, the most recent
approaches rely on large language models, for instance,
\citet{darcy_marg:_2024} use an interconnected ensemble of LLM-based expert agents
to process parts of scientific papers and provide review comments.

\textbf{Pagerank and related methods}

The methods we have discussed so far do not take the bibliographic metadata
(authorship, citations, etc.)\ of communications into consideration, which can be
advantageous in cases where it is unavailable (e.g., in blind reviews), but many
important credibility scoring algorithms rely on this type of data, especially
on the communications' positions in citation networks. In addition to the widely
used, relatively simple closed-form citation-based impact metrics like
Hirsch-index \citep{hirsch_index_2005} or Impact factor \citep{garfield2006history}, which can be used
to establish an author's general credibility in a domain, iterative methods
based on the PageRank link analysis algorithm \citep{brin_anatomy_1998} are highly
influential. PageRank, which was originally used by Google to rank web pages for
their search engine according to the number and quality of links referring to
them has been successfully generalized and applied to several other types of
networks, including ones in which nodes represent publications, institutions or
authors and links correspond to bibliographic references. As
\citet{radicchi_diffusion_2009} demonstrated, this application can be interpreted as an
iterative simulation of the diffusion of author (or institution) credit in the
network. At the start each node has the same amount of credit, but each
iteration redistributes credit proportionally to the weight of local citation
connections, and the process converges to a credit distribution which reflects
the global importance of nodes. Besides being theoretically well-motivated, the
application of PageRank to credit calculation also received some empirical
support, e.g., \citet{dunaiski_evaluating_2016} found that PageRank
variants outperform simpler citation-count-based metrics like Impact factor when
the goal is to find important papers and influential authors.

\textbf{Social networks, trust networks and graph neural
networks}

Decision-making processes often rely on large networks where sets of objects (nodes) are connected through relationships (edges) that all together form a graph-like structure. This is true for social sciences, physical systems and even biology or finance. Keeping the graph information intact during data processing is crucial to preventing losing information \citep{Wu2021}.

Graph Neural Networks (GNNs) provide a state-of-the-art deep learning architecture for processing such input. These neural networks perform graph convolution to propagate features between different nodes and recurrent calculations to incorporate historical data into the node-level representations. To reduce complexity during processing large graphs GNNs often include sampling modules that select a subgraph to perform propagation on. Generating graph, or sub-graph level properties is also possible via pooling modules, which combine information from the affected nodes. Thus, GNNs provide a learnable method for information propagation on the network, plus node-level, edge-level, and graph-level feature vectors that could be further processed \citep{Zhou2020}. GNNs have been adopted to network analysis in the academic domain as well \citep{Iana2021,Liang2022,Zhou2020}. Explicit and explainable credit propagation is still an open question, where model-based deep learning could provide a hybrid solution.

Trust and social networks and their analysis could also provide a viable
solution. A renaissance of network analysis and decision sciences is observable
in the domain of online networks and the credibility of users and information spreading online \citep{Alrubaian2019,Wang2019}. This domain overlaps with
academic network analysis with similar methods throughout both domains
\citep{Kong2019}. Author credit in our system is closely linked to the context
and domain-dependent asymmetric trust metrics. A detailed definition of the trust metric and its propagation is to be determined based on the use case and Board decisions. Propagation methods could include explicit algorithms such as dynamic trust propagation \citep{Sun2023,Jamali2010,Sun2023a}, or data-driven GNNs \citep{Liang2022}.

\subsection{Financial product recommendation}\label{appendix:financial-use-case-ai-background}

Financial products are complex and their advantages and disadvantages may need explanation depending on the specific features of the customer, their goals, and their knowledge level.

Large generative language models could be used as communication agents to interpret financial product parameters in different styles and at different levels. By providing explanations and reasoning over the recommendations, such models facilitate explainability as well. Additionally, logical representations and inference over logical rules could be used to eliminate hallucinations these systems are prone to. This way, an explainable and tractable recommendation system should emerge with the advances in technology. It should be capable of interpreting financial products in common sense tailored to each individual's needs and understanding.

\textbf{Explainable recommendations and
reranking}

It is well known that recommendations with explanations improve the acceptance
rate considerably \citep{Herlocker2000}. The objective of financial product
recommendation is to guide clients towards optimal decisions and rank their
options by considering several key factors. These include (i)~the underlying
causes of being interested in the product type, (ii) the potential outcomes of
the decision, (iii)~the client\textquotesingle s specific goals, (iv)~their
underlying motivations, (v)~their current understanding and awareness, (vi)~their cognitive capabilities, and (vii)~their established, unchangeable
convictions. In addition, decisions should be (viii)~fair and (ix)~transparent.
These constraints form significant obstacles for Assistive AI.

Recommendation systems depend on the field. For example, recommender
systems in the financial sector and streaming services have profound
differences. In the latter case, successful recommendations may be a
great hit leading to the need for similar items. On the other hand, the
customer may seek novelty and similar items may be less successful as
they can be boring. In turn, beyond the nine points listed, the (x)~history and the (xi)~personality trait of the customer can help in
generating recommendations. On the other hand, in the financial sector,
the goal of recommendation is the successful project and not the
novelty.

The number of recommendation algorithms is large. For reviews, see
\citep{Portugal2018} and \citep{Dau2020}. Recommendation of financial products
should take into account several criteria, see (i)--(xi) above and thus the
extensions of general recommender systems, such as context-aware recommender
systems and multi-criteria recommender systems -- see, e.g., \citep{Vu2023} for
Deep learning methods in this field -- can serve the purpose.

Explanation of the recommendation may use statistical arguments and forecasting
subject to the actual conditions. Alternatively, the statistical arguments can
be augmented by explanatory examples, that is success and failure stories.
However, privacy concerns may arise posing an additional challenge for the
recommendation system. In addition, the examples should be congruent in many
aspects with the case of the user. Otherwise, users perceive unfairness and
exclusion that destroy their trust in the recommendation \citep{Chuan}. The
delicate balance between statistical arguments and explanatory examples
targeting feasible low-risk compromises regarding the available financial
products is a great challenge for Assistive AI. Today, explainability is the
major barrier in the deployment of black-box AI methods. The development of more
explainable models and methods of effective explanations started early, see,
e.g., \citep{Gunning2019} and the references therein. XAI has several
interleaved goals, such as trustworthiness, casualty, transferability,
informativeness, confidence, fairness, accessibility, interactivity, and privacy
awareness \citep{Arrieta2020} that will be hard to meet under the constraints
(i)--(xi). A future direction is the analysis of additional data points available
to the bank, such as the information that can be collected from social media or
historical data, e.g., rental payments. Such additional information can be
useful for profiling but raises issues of transparency and accountability due to
unexplained latent variables implicitly generated by the observed cases, and
that may lead to algorithmic bias, too \citep{Chuan}.

Recent information retrieval systems are built on text embeddings: both search
queries and documents to be searched are encoded into vectors representing their
content and retrieval is performed as a nearest-neighbour search in the vector
space. The best-performing systems are based on embeddings produced by
pre-trained LLMs that are adapted specifically to financial information
retrieval. Creating high-quality financial IR data sets and benchmarks is an
active research area \citep{Zhang2020}, in which the usage of LLMs for creating
synthetic documents and queries is also increasing (see, e.g.,
\citealt{Hashemi2023}). In addition to document and query embeddings, vectorized
representations of user-specific information (e.g., recent searches, financial
history etc.)\ are also frequently used for optimal item selection and
ranking (see, e.g., \citealt{Jha2023}).

Today, XAI application tools have model-based approaches, post-hoc explanation
methods and hybrid algorithms. For a recent review, see \citep{Akhai2023}. They
include SHAP, Deep LIFT, ELI5, Anchors, and Skater \citep{Kumar2023}, among
others and the length of the list is increasing \citep{Dwivedi2023}. SHAP and
DeepLIFT try to determine the effects of the input features on the output.
Explain Like I'm 5 (ELI5) and Skater are interpretation tools for ML models. The
technology to comply with the pending legislation is evolving rapidly and could
lead to effective Assistive AI methods in the near future.

Recommendation systems are no exception to this trend. Model factorization,
topic modelling, rule mining and knowledge graph techniques are emerging along
with posthoc techniques such as probing \citep{Zhang2020}. Explainability could
also be important in lately popular multi-stage systems, where the
recommendation or information retrieval process is factorized into multiple
steps. Here the preliminary results are further refined with context-aware
reranking. These reranking steps could be performed by a human (possibly the end-user) \citep{Abdollahi2018} or reranking ML model \citep{Liu2022}. Explainability
plays a dual role here, first, it must provide clues on the "why"-s of the
recommendations if the reranking is performed by a human \citep{Chen2022}.
Second, reranking ML models should also remain explainable by nature, providing
clues on how each ranking decision relates to the preliminary set of
recommendations and the given context. Thus, we can perform multi-step
information retrieval and recommendation over a large set of possibilities with
full explainability \citep{Anand2022}.

Recent studies showcase the up and down-sides of explainability revealing that
due to the easy-to-understand nature of such systems, unfair recommendations are
easier to avoid \citep{Fu2020}, but crafting attacks that poison recommendation
system is also easier to understand as the explanations expose the inner
workings of such systems \citep{Chen2023a}. Mitigating these risks is a matter
of future research and engineering.

An explainable recommendation system for financial products therefore should
apply explainable model definitions, but the recommendation model should only
learn from validated sources (instead of learning from online end-user
interactions such as clicks, likes, or ratings). It is also recommended to
choose an open-ended recommendation interface, where the user can request more
alternatives if needed. The provided recommendations should be paired up with
explaining variables that drove the ranking and reranking models' decisions.
Post-hoc explanations are important to translate the model's short, compact
inner representations into fully-fledged natural language explanations that
satisfy user needs. Counterfactual explanations and historical examples can
provide further explanations to end users as well. Probing and unfairness
measures could be implemented to ensure that a wide enough statistical guarantee
is provided prior to deploying such a system \citep{Anand2022}.

\textbf{Logical rule representation and
inference}

First-order logic (FOL) is a logical framework that could be understood as an
extension of propositional logic, where predicates and quantifiers could be
used. FOL is easy to process in an algorithmic or rule-based manner. On the
other hand, this framework has the representational power of grasping the
meaning of subsets of natural language. Such a subset could be legal documents,
rulings, financial advertisements, declarations \citep{Dragoni2016}.
Translating or parsing these natural language text representations to
first-order logic enables us to perform logical inference, find contradictions
or support regarding the given rules and statements.

Several methods exist for this conversion, Uniform Meaning Representation which
originates from graph-based syntactic and semantic meaning representation
provide a method for converting text into FOL extended with lambda operators
\citep{Gysel2021}. Deep Learning translation models could also be used here, as
well as GPT variants which are specifically trained on FOL conversion tasks
\citep{Han2022,Yang2023}. The performance of these methods is rapidly
increasing, however, there is still room for improvement and further research to
minimize FOL conversion errors, reliable models are still under development.

Adequate FOL-based semantic representation of the content of natural language
sentences make it possible to check their informativeness, consistency, and
to verify the validity of reasoning steps they contain using theorem provers and
model builders such as Isabelle, a generic proof assistant \citep{isabelle2023} and Prover9/Mace4
\citep{McCune20052010}. The fact that first-order logic is undecidable means
that no algorithm is guaranteed to be able to decide these
properties for an arbitrary formula, but, in practice, the strategy of
simultaneously running a theorem prover to try to show the inconsistency (or
un-informativeness) of a formula and a model builder to search for a finite model
of a given size which would establish its consistency (informativeness) proved
to be an effective method in logic-based natural language understanding systems,
and model builders are also useful for constructing explicit discourse models,
from which further information can be extracted \citep{Bos2011}. In cases in
which decidability is a requirement or only representations of a certain type
(e.g., ontology rules) have to be extracted, decidable FOL fragments, e.g.,
description logics can be used (see, e.g., \citealt{Gyawali2017}).

\textbf{Large Language Models in financial product
recommendations}

The finance domain poses several challenges to the application of generic NLP
techniques. While most tasks, e.g., financial product recommendation and
financial question answering have analogues in other domains, its high
complexity, special terminology and time sensitivity have always required
domain-specific solutions. As current state-of-the-art NLP solutions are built
on large language models (LLMs), their domain adaptation requires financial LLMs
that are specifically trained on financial linguistic data. While the first
significant financial LLM, Bloomberg\textquotesingle s proprietary BloombergGPT
\citep{Wu2023}, was trained from scratch on a mixture of domain-specific and
general-purpose texts, most approaches start with general LLMs and fine-tune
them on relatively small but carefully assembled financial datasets, with
frequent emphasis on examples of financial instruction following and assistance
(see, e.g., \citep{Chen2023,Xie2023}). While both approaches result in
domain-adapted financial LLMs that outperform their general counterparts by a
significant margin, fine-tuning, especially with parameter-efficient methods
(PEFT) achieves competitive results far more cost-effectively (see, e.g.,
\citealt{Xie2023}), and makes it viable to update the models frequently, which
is essential in highly time-sensitive applications.

Large Language Models (LLMs) experience a wide range of research concerning
coupling them with external tools. Here the LLM is the main interpreter, or
interface that contacts the end user, while it also interacts with tools
available through APIs \citep{Yang2023a}. These tools could include program
calls, external applications, web sources, information retrieval pipelines, or
even other LLMs with different configurations. Using tools can improve
task-specific performance and user experience as the tools only need to be
active when a user request arises \citep{Qin2023}.

In our financial product recommendation these tools could be:
\begin{itemize}
\item
  Customer information retrieval from the database, if authorized;
\item
  Task-oriented LLM instance for filling in application forms;
\item
  Recommendation system call;
\item
  Retrieval augmented generation system for detailed financial product
  information.
\end{itemize}

This way the system will be able to retrieve related information if needed
\citep{Gao2022}. It can enter a form-filling process and can also call for the
recommendation system if needed, while it can provide detailed information by
accessing that financial product's description which is to be explained. The
scheduling of the tool calls depends on the core LLM that interprets the user
interactions according to the conversation script.

\subsection{Financial credit scoring}
\label{appendix:financial_credit_scoring}

The domain of financial credit scoring is the same as that of the financial product recommendation. The notable difference is in the targeted users. Here, the AI system should assist not only the customers but also the employees and decision-makers of a financial institution. Similarly to the previous case, the process of scoring should remain logically grounded and explainable. However, logical inference is somewhat different as it should include the analysis and detection of inconsistent or contradicting rules in the knowledge base. Thus non-compliant processes could be detected and improved.

As the AI methods of financial credit scoring and financial product recommendation are very similar, we refer the reader to Sect.~\ref{appendix:financial-use-case-ai-background}.

\end{document}